\documentclass[conference]{IEEEtran}
\IEEEoverridecommandlockouts

\usepackage{mathptmx} %
\usepackage{fancyhdr}
\usepackage[normalem]{ulem}
\usepackage{flushend}
\usepackage{comment}
\usepackage{listings}
\usepackage{color}
\usepackage{threeparttable}
\usepackage{booktabs}
\usepackage{tabularx}
\definecolor{vgreen}{RGB}{104,180,104}
\definecolor{vblue}{RGB}{49,49,255}
\definecolor{vorange}{RGB}{255,143,102}
\usepackage{algorithm}
\usepackage{algorithmicx}
\usepackage{algpseudocode}
\usepackage{mathtools}
\usepackage{pifont}
\usepackage{footnote}
\usepackage[compatibility=false]{caption}
\usepackage{subcaption}
\usepackage{paralist}
\usepackage{multirow}
\usepackage{cite}
\usepackage{hyperref}
\usepackage[nameinlink,capitalize]{cleveref}

\definecolor{vgreen}{RGB}{104,180,104}
\definecolor{vblue}{RGB}{49,49,255}
\definecolor{vorange}{RGB}{255,143,102}

\hypersetup{
  bookmarks=true,
  breaklinks=true,
  letterpaper=true,
  colorlinks,
  linkcolor=black,
  citecolor=blue,
  urlcolor=black
}

\newcommand\longvar[1]{\mathchardef\UrlBreakPenalty=100
\mathchardef\UrlBigBreakPenalty=100\url{#1}}

\def\shortauthors{M. Tarek Ibn Ziad, M. Arroyo, E. Manzhosov, V. P. Kemerlis, and S. Sethumadhavan}

\fancypagestyle{plain}{%
  \fancyhf{}

  \fancyhead[R]{\sffamily\footnotesize\shortauthors}
  \fancyfoot[C]{\thepage}
}

\def\pname{PNS}
\def\pnameEx{Phantom Name System}

\def\origNO{Original}
\def\phanNO{Phantom}
\def\ENCP{\texttt{ENCP}}
\def\ENCPex{\texttt{ENCP RegX}}
\def\DECP{\texttt{DECP}}
\def\DECPex{\texttt{DECP RegX}}
\def\ShadowStackEx{\texttt{Secret Domain Stack}}
\def\ShadowStackNormal{Secret Domain Stack}
\def\ShadowStackShort{SDS}
\def\DomainShift{$\Delta$}
\def\SecurityShift{$\delta$}
\def\PTRENCNormal{Lightweight Pointer Encryption}

\def\PTRENCShort{PtrEnc}
\def\TRAP{\texttt{TRAP}}
\def\BBL{BBL}
\def\BBLEx{basic block}

\def\ze{they}

\def\ToolchainBaseline{Baseline}
\def\ToolchainPasNoTrap{\pname{}}
\def\ToolchainCcfilite{\PTRENCShort{}Lite}
\def\ToolchainCcfifull{\PTRENCShort{}Full}
\def\ToolchainPasTrap{\pname{}-\texttt{TRAP}}
\def\ToolchainPasccfi{\ToolchainPasNoTrap{}-\ToolchainCcfilite{}}

\def\ToolchainSwisoEx{Naive Name Confusion}
\def\ToolchainSwiso{NNC}

\crefformat{section}{#2\S~#1#3}
\crefformat{subsection}{#2\S~#1#3}
\crefformat{subsubsection}{#2\S~#1#3}
\crefformat{figure}{#2Fig.~#1#3}
\crefformat{table}{#2Tbl.~#1#3}

\newcommand{\fakesubsub}[1]{\noindent\textbf{#1}}

\makeatletter
\newcommand{\AND}{%
\end{@IEEEauthorhalign}
\hfill\mbox{}\\[1.5\baselineskip]
\mbox{}\hfill\begin{@IEEEauthorhalign}
}
\makeatother

\begin{document}

\title{Using Name Confusion to Enhance Security}
\author{
  \IEEEauthorblockN{M. Tarek Ibn Ziad}
  \IEEEauthorblockA{Columbia University}
  \textit{mtarek@cs.columbia.edu}
  \and
  \IEEEauthorblockN{Miguel A. Arroyo}
  \IEEEauthorblockA{Columbia University}
  \textit{miguel@cs.columbia.edu}
  \and
  \IEEEauthorblockN{Evgeny Manzhosov}
  \IEEEauthorblockA{Columbia University}
  \textit{evgeny@cs.columbia.edu}
  \AND
  \IEEEauthorblockN{Vasileios P. Kemerlis}
  \IEEEauthorblockA{Brown University}
  \textit{vpk@cs.brown.edu}
  \and
  \IEEEauthorblockN{Simha Sethumadhavan}
  \IEEEauthorblockA{Columbia University}
  \textit{simha@cs.columbia.edu}
}

\maketitle

\begin{abstract}

We introduce a novel concept, called Name Confusion, and
demonstrate how it can be employed to thwart multiple classes of code-reuse
attacks. By building upon Name Confusion, we derive \pnameEx{} (\pname{}): a
security protocol that provides multiple names (addresses) to program
instructions. Unlike the conventional model of virtual memory with a one-to-one
mapping between instructions and virtual memory addresses, \pname{} creates N
mappings for the same instruction, and randomly switches between them at
runtime. \pname{} achieves fast randomization, at the granularity of basic
blocks, which mitigates a class of attacks known as (just-in-time) code-reuse.

If an attacker uses a memory safety-related vulnerability to cause any of the
instruction addresses to be different from the one chosen during a fetch, the
exploited program will crash. We quantitatively evaluate how \pname{} mitigates
real-world code-reuse attacks by reducing the success probability of typical
exploits to approximately $10^{-12}$. We implement \pname{} and validate it by running
SPEC CPU2017 benchmark suite. We further verify its practicality by adding it to 
a RISC-V core on an FPGA. Lastly, \pname{} is mainly designed for resource
constrained (wimpy) devices and has negligible performance overhead, compared to
commercially-available, state-of-the-art, hardware-based protections.

\end{abstract}

\section{Introduction}\label{sec:introduction}

Virtual memory addresses serve as references, or \emph{names}, to objects
(i.e.,~instructions, data) during computation. For instance, every instruction
in a program is uniquely identified (at run time) with a virtual memory
address: the value in the Program Counter (\texttt{PC}). Typically, the virtual memory
address assigned to an instruction is kept constant and unique for the life
time of the program. In this work, we show that having multiple names for an
instruction---at any given time instant---improves the security of the
system with minimal hardware support without performance degradation.

How can having multiple names improve security? Given multiple names for an
instruction, we define a \emph{security protocol} that specifies a random
sequence of names to be used during execution. If the attacker does not follow the
security protocol by supplying an incorrect name, the exploited program will
crash. In other words, if there are~$N$
addresses (names) per instruction, and if the attacker has to reuse~$P$
instruction sequences to complete an attack, the probability of detecting the
attack is~$1 - (1/N)^P$, without any false positives. For example, for $N=256$
and $P=5$, then the probability of an attack succeeding is $1$ in $1$ trillion.
This kind of protection makes this technique suitable to be used as a standalone
solution, or in tandem with other, heavier-weight hardening mechanisms. We refer
to such classes of architectures as \emph{Name Confusion Architectures}.

Name confusion is fundamentally different from other hardening paradigms. For
example, in the information-hiding paradigm~\cite{goktas2016:InfoHiding},
the program addresses (or parts of them) are kept a secret, but there is
only one name per instruction. Similarly, Instruction Set Randomization (ISR)
techniques~\cite{Gaurav2003:ISR,Antonis2013:ASIST,Sinha2017:ISR} randomize the
encoding of instructions in memory, while also maintaining a unique instruction
name per program execution. In the metadata-based paradigm, such as
Control-Flow Integrity (CFI)~\cite{Abadi2005,burow2017control}, the set of
targets (names) that can result from the execution of certain instructions
(i.e.,~indirect branches) are computed statically and checked during execution.
In moving target paradigms, such as Shuffler~\cite{David2016:shuffler} and
Morpheus~\cite{Gallagher2019:Morpheus}, the names of instructions change over
time; however, at any given time, there is only one valid name/address for an
instruction.

In this work, we explore an application of a name confusion-based architecture,
and show how it is used to mitigate a class of attacks known as code-reuse
(ROP~\cite{Shacham2007,Buchanan2008,Checkoway2010}, JOP~\cite{Bletsch2011:JOP},
COP~\cite{goktas2014out}), including their just-in-time
variants~\cite{snow2013}. The instance we consider, called\emph{~\pnameEx{}}
(\pname{}), provides up to~$N$ different names, for any instruction, at any
given time, where~$N$ is a configurable parameter (it is set to~$256$ in our
design). The security protocol for~\pname{} is simply a truly random selection
among the different names. Specifically,~\pname{} works as follows: during
instruction fetch, the address used to fetch the instruction is randomly chosen
from one of the~$N$ possible names for the instruction, and the instruction is
retrieved from that address. From that point on, any PC-relative addresses used
by the program relies on the name obtained during fetch. If the attacker's
strategy causes any of the PC-relative addresses to be different from the one
used during the fetch, then an invalid instruction will be executed, leading to
unexpected effects, such as an alignment, or instruction-decoding, exception.
These unexpected effects lead to program crashes that can work as signals of bad
actions taking place especially in the case of repeated crashes. Orthogonal
mechanisms that turn these signals into a defensive advantage
exist~\cite{Locasto2005:AC}.

A naive implementation of~\pname{} would require each instruction to be stored
in~$N$ locations so they will have~$N$ names. Consequently, the capacity of all
PC-indexed microarchitectural structures would be divided by~$N$, heavily
impacting performance. Further, this requires changes to the compiler, linker,
loader, \emph{etc.} In this work, we use a simple technique to avoid these
problems: we intentionally \emph{alias} the different instruction
names/addresses so they point to the same instruction, allowing us to serve
the~$N$ instructions from one copy. This idea is similar to how multiple
virtual addresses can point to the same physical addresses (used to implement
copy-on-write~\cite{Bovet:2002:ULK}) with two key differences: first,
in~\pname{} the~$N$ names correspond to the same virtual address, not a
physical address; and second, the~\pname{} addresses do not need to be page-aligned
as required for data synonyms---i.e.,~\pname{} virtual address names can be
arbitrarily offset. The first difference ensures that~\pname{} can be
handled at the application level without requiring significant changes to the
operating system (OS), which manages the virtual-to-physical address mappings,
while the second is key to providing security.

With the above optimizations, we show that ROP/JOP/COP attack protection is 
provided at almost no performance overhead and without any binary changes. We
further show, that our scheme can be combined with previously known
techniques~\cite{Cowan2003:PointGuard, Tuck2004:PointerEncryption,
Mashtizadeh2015, Hans2018} that encrypt instruction addresses stored in the
heap or the global data section(s), viz., function pointers, to provide robust
security against even larger class of attacks, such as
COOP~\cite{Schuster2015}. The combined protection scheme has~$6\%$ performance
overhead, making it better than state-of-the-art commodity security solutions,
like the ARM pointer authentication code (PAC)~\cite{Qualcomm2017} that is available in the latest iPhone
devices, and has the additional benefit of not requiring a~$64$-bit
architecture. Supporting non-$64$-bit architectures is important as they make
up the majority of the computing devices that exist nowadays: in $2018$,
$11.75$ million servers shipped worldwide~\cite{Statista:Servers} vs.  $28.1$
billion $32$-bit (or smaller) microcontrollers~\cite{Statista:MCU}.

\section{Name Confusion Architecture}\label{sec:systemoverview}

A name confusion architecture assigns different addresses, or names, to any
contiguous group of instructions randomly at runtime. In this section, we
introduce \pname{}, a security protocol derived from the principles of name
confusion architectures. 

\pname{} consists of~$N$ \emph{phantoms} (domains). It
requires every instruction in the program to have $N$ unique names. To assign
the names, we use a mapping function,~$name_p = f(va, p)$, which takes the
instruction virtual address,~$va$, and a phantom index,~$p$ as inputs and
returns the phantom name, $name_p$. This way any instruction is mapped
by~$f$ to unique location in each of the~$N$ phantoms. The function $f$ does not
have to be kept a secret, as security is purely derived from the random
selection of $p$ at fetch time. For mapping a phantom name
to its original virtual address, we use the inverse function,~$va =
f^{-1}(name_p)$. To enable the inverse function we ensure that the phantom name
encodes the phantom index,~$p$ as part of the phantom address.

\begin{figure}[!t]
  \centering
  \includegraphics[width=0.45\textwidth]{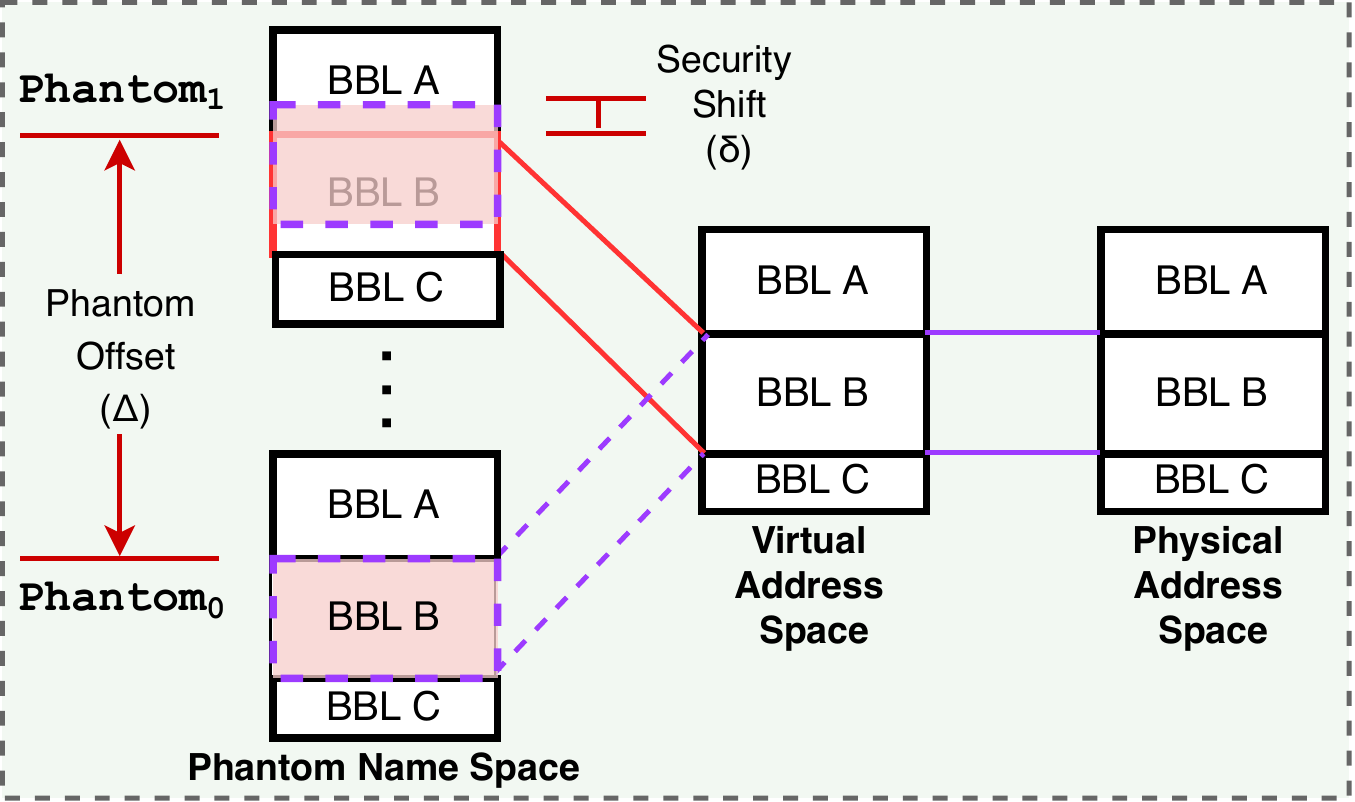}
  \caption{Basic block mapping for~\pname{}. BBLs are only duplicated in the phantom space. } \label{fig:IsoVM}
\end{figure}

\subsection{\pname{} Framework} \label{sub:pas-frame}

There are four main operations to realize~\pname{}:
\emph{Populate}, \emph{Randomize}, \emph{Resolve}, and \emph{Conceal}.

\fakesubsub{Populate.}  \pname{}
creates multiple phantoms of basic blocks, and populates
them in the phantom name space. The left-hand side
of Figure~\ref{fig:IsoVM} shows a program with two~\phanNO{}s, such that every
basic block (BBL) has two different names (addresses) in~\phanNO{}$_0$ (aka the
original domain) and~\phanNO{}$_1$.~\pname{} separates the two~\phanNO{}s
by a phantom offset,~\DomainShift{}, in the phantom space. To add discrepancy
between the~\phanNO{} copies, we introduce a
minor security shift,~\SecurityShift{}, so that they are not perfectly
overlapped after removing~\DomainShift{}. This is shown by the shaded basic
block in Figure~\ref{fig:IsoVM} and is necessary for security, as will be illustrated in Section~\ref{sub:cra_pro}.
The inverse mapping
function~$f^{-1}$ maps all phantoms to a single name in the virtual address space,
which is then translated to a physical address by the OS.

\fakesubsub{Randomize.} We modify the
hardware to randomize program execution between the~\phanNO{}s at runtime.
For example, some basic blocks will be executed from~\origNO{} (\phanNO{}$_0$) while
other basic blocks will be executed from any other~\phanNO{}. Correctness is
unaltered because all~\phanNO{}s provide the same functionality
by construction.

\fakesubsub{Resolve.} Accessing different instruction names
at runtime incurs additional performance overheads as each name
needs to be translated to a virtual address and then a physical one before usage. To mitigate
this problem,~\pname{} uses the inverse mapping function~$f^{-1}$
to resolve the different~\phanNO{}s to their archetype basic block.
By doing so, the processor back-end continues to operate as if there
is only one copy of the program in the phantom name space.

\fakesubsub{Conceal.} Normal programs push return addresses to the architectural stack
to help return from non-leaf function calls. The attacker may learn the
domain of execution, the~\phanNO{} index, by monitoring the
stack contents at runtime using arbitrary memory disclosure
vulnerabilities~\cite{snow2013}.
Thus to preserve name confusion, we need to conceal the execution domain of
the instructions.

\subsection{\pname{} Construction} \label{sub:pas-construction}
In this section, we discuss alternative
design choices for the different operations in the~\pname{} framework.

\fakesubsub{Populate.} Many approaches can be used to populate the \phanNO{}s.
One approach is to use the most significant bits (MSBs)
to separate the program copies in the phantom
space. For example, a~\DomainShift{} of~\texttt{0x8000\_0000\_0000\_0000}
will create two phantoms on~$64$-bit systems, where each phantom resides in one half of the address space. This approach
is acceptable for~$64$-bit systems
because \texttt{VA} allows for~$64$ bits, yet only~$48$ are used in
practice, leaving the higher order bits available for phantom addresses.
However, this is costly
for~$32$-bit systems as it will reduce the effective range of
addresses a program can use by half.
Instead, to store the phantom index we add $n$ additional bits to the hardware program counter,
while maintaining the~$32$-bit virtual address space of the program. This allows \pname{} to generate
~$N = 2^n$ phantoms. Specifically,~$f$,
sets the additional~$n$ bits at control-flow
transitions to randomize the execution at runtime.
For simplicity, we set the phantom offset as \DomainShift{}~$ = 1 \ll 32$ and
the minor security shift of any
phantom to be a multiple of the phantom index (i.e.,~\SecurityShift{}$_p = p \times
$\SecurityShift{}). We elaborate more
on the~\pname{} realization in Section~\ref{sec:microarchitecture}.

\fakesubsub{Randomize.}  \pname{} can randomize program addresses
at any level of granularity, ranging from individual instructions to entire programs. In the rest
of the paper, we use basic blocks as our elements of interest. We do not
evaluate finer granularities here due to the lack of a strong security need. We
define the basic block as a single entry, single exit region
of code. Thus, any instruction that changes the \texttt{PC} register
(referred to by control-flow instructions, such as \texttt{jmp,
call, ret}) terminates a BBL and starts a new one.\footnote{Some
compilers, such as LLVM, deviate from this definition and treat
\texttt{call} instructions as part of the BBL.}

\fakesubsub{Conceal.} We can prevent attackers from learning the execution
domain in a number of ways. One straightforward way is to encrypt the return address
with a secret key and only decrypt it upon function return.
Another key-less, and low overhead, method that we implement
is to split this information so that the public part is what is common between
the phantom domains, and the private part that distinguishes the domains is
hidden away without architectural access.

We split the return addresses between the architectural stack and
a new hardware structure called
the~\ShadowStackEx{} (\ShadowStackShort{}),
which by construction is immutable to external writes. \ShadowStackShort{}
achieves this goal by splitting the return address~$(32 + n)$ bits into two parts;
the~$n$-bits, which represent the \emph{phantom index} ($p$), and the lower~$32$ bits
of the address, which encodes the \emph{security shift} (\SecurityShift{}).
With each function \texttt{call} instruction, the
lower~$32$ bits of the return address are pushed to the architectural
(software) stack, whereas the phantom index~$p$ is pushed onto the~\ShadowStackShort{}.
A \texttt{ret} instruction pops the most recent~$p$ from the
top of~\ShadowStackShort{} and concatenates it with the return
address stored on the architectural stack in memory. While under attack,
the return address on the architectural stack will be corrupted by the attacker.
However, the attacker cannot access~\ShadowStackShort{} so they cannot reliably
adjust the malicious return address to correctly encode~\SecurityShift{},
leading to an incorrect target address after~\pname{}
merges the malicious return address with the phantom index~$p$ from~\ShadowStackShort{}.
Deployment issues with the~\ShadowStackShort{} such as sizing, overflows,
multithreading, \emph{etc.} are described in Section~\ref{sec:deployment}.

\subsection{\pname{} Correctness} \label{sub:corr}

Since the addresses are selected during a fetch, how can we be assured
that all PC relative computations are used in the correct way as
encoded in the original program?  In order to discuss the correctness
of \pname{} construction and operation, we consider the structured
programming theorem~\cite{Dahl1972:StructuredProg}, in which a
program is composed from any subset of the control structures:
\textit{Sequence}, \textit{Selection}, \textit{Iteration}, and
\textit{Recursion}.  We show that \pname{} does not affect the four
structures. For simplicity, let us assume~$n = 1$, so that only two phantoms
exist,~\origNO{} or~\phanNO{}. First, the \textit{Sequence} structure represents a
series of ordered statements or subroutines executed in sequence.
\pname{} guarantees this property by executing the statements
(instructions) of a BBL in the same domain of execution (either~\origNO{}
or~\phanNO{}).

For handling the \textit{Selection} structure, let us assume that
a program is represented by a binary tree with a branching factor
of~$2$. Hence, we define two types of such trees. Type~\texttt{I}
is the tree where nodes are represented by the BBLs of \emph{committed}
instructions. Type~\texttt{II} is the tree where each node has the
address of the first instruction in the \emph{executed} BBL. The
edges are given by the direction taken by the last instruction of
each BBL. The root of the tree is the first instruction fetched
from the \texttt{\_start()} section of a binary (or the address of this
instruction in the address based tree). The leaf node is the last
BBL of the program (or the address of this BBL in trees of
Type~\texttt{II}). In the case of \pname{}, every taken branch on the
tree of Type~\texttt{I} is the same as every taken branch on the
tree of Type~\texttt{II}, i.e., program functional decisions are
not affected. However, the contents of the tree nodes in
Type~\texttt{II} trees would be different for each program execution,
as each BBL will be fetched from either~\origNO{} or~\phanNO{}
domain and addresses of these domains differ by~\DomainShift{}. In
other words, after the branch is resolved to be taken or not
taken,~\pname{} operates on the outcome and randomly chooses the
domain of execution for the next basic block.

The above argument also applies for the \textit{Iteration}
control-structure, in which the same basic block is executed multiple
times. \pname{} does not change the functionality of the basic block,
however, each time the basic block is executed it will be executed in
one of the phantom domains. The \textit{Recursion} construct is
similar to iterative loops and thus is guaranteed to be executed
correctly with~\pname{}. The same proof holds for~$n > 1$ by making
the branching factor of the tree equals~$2^n$. From the user perspective,
the program will produce the same result, since the order and flow of
instructions has not
changed (Type~\texttt{I} trees are unique). However, from the perspective
of an entity that observes only addresses, each execution of a
program appears to be a different sequence of
addresses since the addresses of individual basic blocks
will be altered (Type~\texttt{II} trees are not unique). This enables
a substantial security benefit as we show next.
\section{Hardware Design}\label{sec:microarchitecture}

\begin{figure}[!t]
  \centering
  \includegraphics[width=0.45\textwidth]{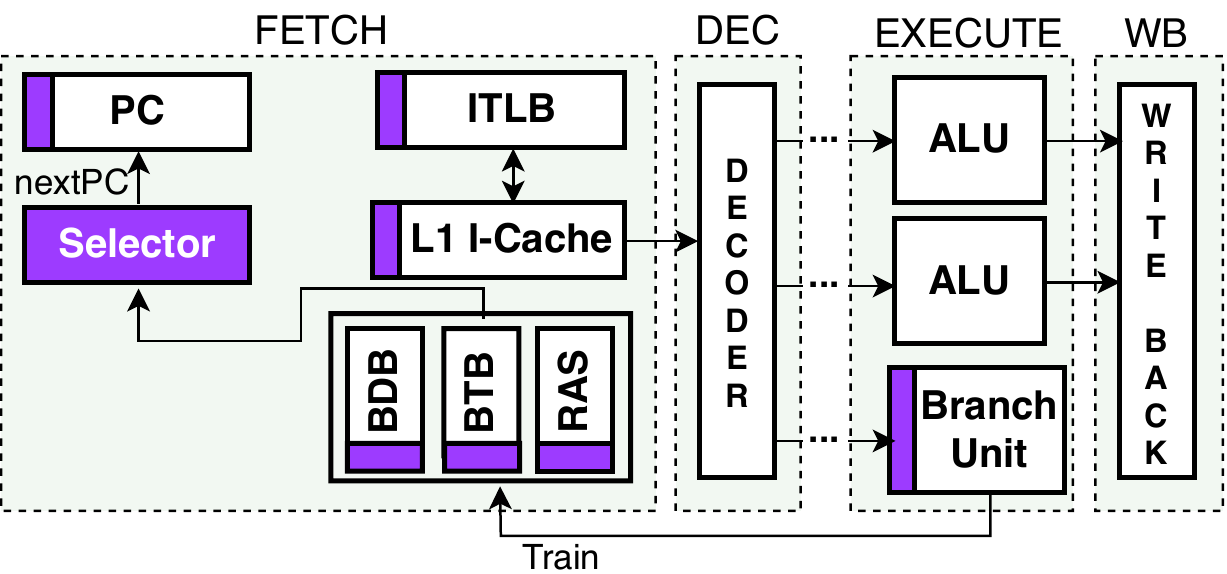}
  \caption{Processor pipeline with \pname{} hardware.}\label{fig:PASuarch}
\end{figure}

Figure~\ref{fig:PASuarch} summarizes our modifications to
support \pname{}.  The changes are limited to structures that operate
on \texttt{PC}.

\subsection{Selector} \label{sub:div-micro}

With~\pname{}, each \texttt{PC} is extended by (additional)~$n$-bits, 
dubbed the \emph{phantom index} ($p$). So, a program counter from phantom~$p$ 
will have the following format: 

\begin{equation}
\label{eqn:pc}
\texttt{PC$_p$}[31+n:0] = \{ p[n-1:0] , \texttt{PC}[31:0] \}   
\end{equation}

The Selector~($S$) is responsible for adjusting the \texttt{PC}
before executing any new BBL so that the execution flow cannot be predicted 
by the attacker. Specifically, the selector takes the predicted target for
a branch (\texttt{PC$_{new}$}) with control-flow signal~$s$ as
input:~$s$ is set to one if the Branch Predictor Unit (BPU) has
a predicted target for this instruction, or to zero otherwise. The selector
generates the~\texttt{nextPC} as the output. If~$s$ equals one, the selector
generates an~$n$-bit random phantom index~$p_{next}.$\footnote{This can be implemented
using~$n$ metastable flip-flops~\cite{Kim90:meta}.} Based 
on~$p_{next}$, the selector adjusts
the~\texttt{nextPC} according to Equation~\ref{eqn:pcAdj}.

\setlength{\belowdisplayskip}{0pt} \setlength{\belowdisplayshortskip}{0pt}
\setlength{\abovedisplayskip}{0pt} \setlength{\abovedisplayshortskip}{0pt}
\begin{equation}
\label{eqn:pcAdj}
\begin{split}
\texttt{nextPC}[31+n:0] = & \{ p_{next}[n-1:0] , \\  
                                            &  \texttt{PC$_{new}$} - (p_{next} - p_{new}) \times \delta \}
\end{split}
\end{equation}

\noindent Note that~$p_{new}$ is the phantom index of the predicted target 
\texttt{PC$_{new}$}. For example, assuming~$n = 8$-bits, we have~$2^8 
= 256$ phantoms. If \texttt{PC$_{new}$} corresponds to the fifth phantom (i.e., $p_{new} 
= 5$) and the selector randomly chooses the eighth phantom (i.e., $p_{next} = 8$), 
\texttt{nextPC} will equal~$\{ 8,  \texttt{PC$_{new}$} - 3 \delta\}$. On the 
other hand, if the selector randomly chooses the second phantom (i.e., $p_{next} = 2$), 
\texttt{nextPC} will equal~$\{ 2,  \texttt{PC$_{new}$} + 3 \delta\}$. As the 
security shift~\SecurityShift{} is only used to break the overlapping between 
the names in different phantoms, it can be arbitrarily set to a single byte on CISC 
architectures or multiples of the instruction size on RISC architectures.

\fakesubsub{Performance Optimization~$\#1$.}
The aforementioned selector
adds one cycle latency to the~\texttt{nextPC}
calculations in the fetch stage. To alleviate this,
\emph{we move the selector to the commit stage}---placing the selector
at the commit stage allows us to mask latency overheads needed for
target address adjustments so that it does not affect performance.

At the commit stage, the target of the branch instruction is known and sent back
to the fetch stage to update (\texttt{train}) the BPU buffers.
At this point, the selector will adjust the target address by
using~$p_{next}$, as explained above and update the BPU buffers
with~\texttt{nextPC}.  This ensures that the next execution for
this control-flow instruction will be random and unpredicted.
To bootstrap the first execution of a control-flow instruction, 
we consider the two possible cases: correct and incorrect prediction.
If the first occurrence of the control-flow
instruction is correctly predicted to be~\texttt{PC + 4} (falling
through), then the selector will keep using the current domain of
execution (unknown to the attacker) for the next BBL. If the first
occurrence of the control-flow instruction is incorrectly predicted,
it would be detected later on in the commit stage and the pipeline
will be flushed. In this case, the selector will adjust the resolved
target address by using~$p$ (unknown to the attacker) and update the BPU buffers
with~\texttt{nextPC}.

\subsection{Branch Prediction Unit (BPU)} \label{sub:BPU}

The branch prediction unit stores a record of previous
target addresses in the branch target buffer (BTB), and the recent
return addresses in the return address stack (RAS). For the
current~\texttt{PC} value, the BPU checks if the corresponding entry
exists in the BTB by indexing with the PC. If it exists, the found target
address becomes the~\texttt{nextPC}. Otherwise, \texttt{nextPC}
is incremented to~\texttt{PC + 4} (or \texttt{PC + Instruction
size}). If the predicted target address turns out to be incorrect
later in the instruction pipeline, the processor re-fetches the
instruction with the correct target address (available usually at
the execute stage of the branch instruction) and nullifies the
instructions fetched with the predicted target address.

\begin{figure}[!t]
  \centering
  \includegraphics[width=0.45\textwidth]{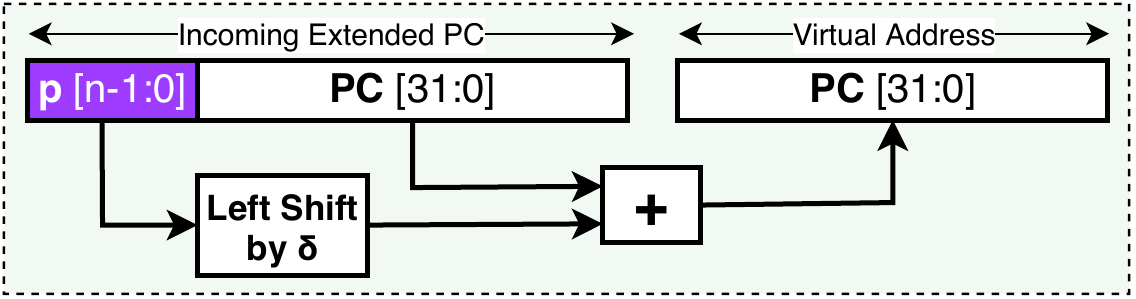}
  \caption{Mapping the extended PC (i.e.,~the phantom name) to the virtual address before indexing into the micro\-architectural structures.}\label{fig:PASmask}
\end{figure}

\fakesubsub{Performance Optimization~$\#2$}.~\pname{}
assigns~$N$ different addresses for the same control-flow instruction.
In this case, we will have multiple entries in the prediction tables
for the same effective instruction; this reduces the capacity
to~$\frac{100}{N}\%$. To handle this issue, \emph{we map the incoming 
phantom address to its original name before indexing into the BPU tables}, 
as shown in Figure~\ref{fig:PASmask}. We do so by modifying the hashing 
function of the BPU tables to avoid adding any latency to the lookup operation. 
This way we guarantee that all phantom addresses (names)
map to the same table entry. After indexing, we get the desired
values from the prediction tables. As explained in
Section~\ref{sub:div-micro}, the \texttt{nextPC} values stored in
the BTB are already chosen at random from the last successful commit
of this control-flow instruction (or any of its phantoms). The branch direction prediction
results (Taken vs. Not Taken) in the branch direction buffer (BDB)
remain the same.

\subsection{Translation Look-aside Buffer (TLB)} \label{sub:tlb}

\fakesubsub{Performance Optimization~$\#3$.} Similar to the BPU
buffers, the fact that we have~$N$ variants of every BBL with 
different virtual addresses may lead to multiple different
virtual-to-physical address entries in the TLB for the same 
translation, reducing its capacity to~$\frac{100}{N}\%$. To avoid
potential performance degradation, \emph{we map the incoming 
phantom address to its original name before accessing the ITLB}. 
For example, the following two phantom 
addresses,~\texttt{\{2, 0x00BB\_FFF4\}} and~\texttt{\{0, 0x00BB\_FFF8\}}, will 
point to the same virtual address,~\texttt{0x00BB\_FFF8}. 
This common virtual address has a unique mapping to a physical 
address,~\texttt{0x0011\_DDFC}, that is stored in the ITLB.
Thus, the translations 
related to all~\phanNO{}s map to a single entry in the ITLB, while we do not 
modify physical addresses so that the stored physical address
part of the translation remains unaffected.

\subsection{Instruction Cache} \label{sub-cache}

\fakesubsub{Performance Optimization~$\#4$.}
Creating~$N$ variants of the
code sections for each program means that the L1-I\$ capacity would
be effectively reduced to~$\frac{100}{N}\%$.
\pname{} maps the incoming phantom address to its virtual address 
before accessing the L1-I\$ (in case of virtually-indexed caches) or performing 
the tag comparison (in case of virtually-tagged caches).\footnote{
No changes are needed for Physically-Indexed Physically-Tagged (PIPT) caches.}
This represents our simple inverse mapping 
function,~$f^{-1}$. The latency of the adjustment operations (shifting and addition) 
can be masked within the cache read operation. 
This incoming address adjustment ensures that while
executing a BBL$_{\phanNO{}}$ we fetch the correct instruction.

\subsection{Execution Unit} \label{sub:exe-unit}

\fakesubsub{Performance Optimization~$\#5$.} If the target architecture 
allows forwarding the \texttt{PC} register through the pipeline for regular 
instructions, we make sure that \emph{the \texttt{PC} register is always 
mapped to the virtual address before operating on it}. This mapping may 
introduce additional latency for the execute stage as it should be done 
\textit{before/after} it.
To mask such latencies, one solution 
is to always forward the two versions, \phanNO{}$_p$ and \origNO{}, of the \texttt{PC} 
register to the desired execution units. Although such 
a solution completely hides the adjustment latency, it may increase the 
execution unit(s) area.

\subsection{\ShadowStackNormal{}} \label{sub:shadowstack}

\fakesubsub{Performance Optimization~$\#6$.}  Unlike prior work,
which stores a complete version of the return addresses (e.g., $32$-bit on \texttt{AARCH32})
in what is called a shadow stack~\cite{Burow2019:shadow}, we only store~$n = 8$ 
bits per return address. To minimize silicon area within the processor and
facilitate managing the~\ShadowStackShort{}, as discussed in
Section~\ref{sub:pas-construction}, we do not need to store the full return
address. This structure does not introduce additional latency as it is accessed
in parallel to the normal architectural stack access. We evaluate the optimal
size of~\ShadowStackShort{} in Section~\ref{sec:evaluation}.

\section{Code Reuse Protection with \pname{}} \label{subsec:pas-cra}

In this section, we summarize code-reuse attacks and defenses, 
and discuss how~\pname{} is used to mitigate such attacks.   

\begin{figure*}[!th] 
  \centering
  \includegraphics[width=0.9\textwidth]{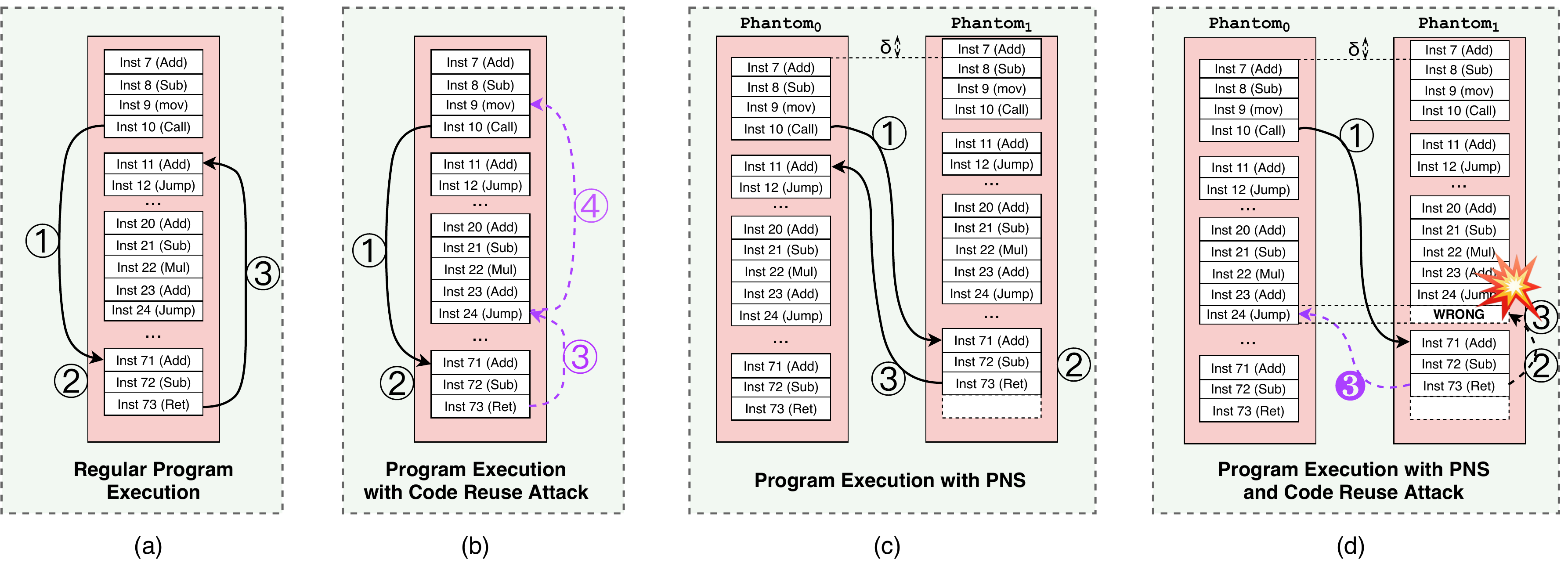}
  \caption{\pname{} effect on CRAs: (a)~shows the regular program execution;
(b)~shows successful CRA via ROP; (c)~shows regular program execution with \pname{}; (d)~shows ROP failure with \pname{} due to missing the desired gadget.}
  \label{fig:Iso}
\end{figure*}

\subsection{Background}\label{sec:background}

Attacks that chain together gadgets
whose last instruction is a \texttt{ret} are known as return oriented
programming (ROP) attacks~\cite{Shacham2007,Buchanan2008}. To mount a ROP
attack, the attacker has to first analyze the code to identify the respective
gadgets, which are sequences of instructions in the victim program (including
any linked-in libraries) that end with a return. Second, the attacker uses a
memory corruption vulnerability to inject a sequence of return addresses
corresponding to a sequence of gadgets. When the function returns, it returns
to the location of the first gadget. As that gadget terminates with a return,
the return address is that of the next gadget, and so on. As ROP executes
legitimate instructions belonging to the program, it is not prevented by
W\textasciicircum{X}~\cite{drepper2004security}. Note that variants of 
ROP that use indirect \texttt{jmp}
or \texttt{call} instructions, instead of \texttt{ret}, to chain the execution
of small instruction sequences together also exist, dubbed jump-oriented
programming (JOP)~\cite{Bletsch2011:JOP} and call-oriented programming
(COP)~\cite{goktas2014out}, respectively. 

\subsection{Currently Deployed Mitigations} \label{subsec:current-mitigations}

The standard mitigation technique against ROP/JOP/COP, and pretty
much every Code Reuse Attack (CRA) variant, is address space layout randomization (ASLR),
which is currently a well-adopted defense, enabled on (pretty much)
every contemporary OS~\cite{Ollie2007}.  Essentially, ASLR forces
the attacker to first \emph{disclose} the code layout (e.g., via a
code pointer) to determine the addresses of gadgets.  Snow \emph{et
al.}~\cite{snow2013} observed that typical programs have multiple
memory disclosure vulnerabilities.  They developed a just-in-time
ROP (JIT-ROP) compiler that explores the program's memory, disassembling
any code it finds (in memory), as well as, searching for API/system
calls.  Then, they construct a compatible code-reuse payload
\emph{on the fly}.  Note that, in principle, JIT-ROP is not restricted
to dynamically stitching together only ROP payloads; it can also
compile JOP, COP, or any other code-reuse payload.

Recently, ARM introduced PAC
in \texttt{Armv8.3A}, which is 
implemented in the Apple's iPhone XS SoC~\cite{Qualcomm2017}. The
idea is based on a concept known as cryptographic control-flow
integrity (CCFI)~\cite{Mashtizadeh2015}. For every code pointer,
such as return addresses and function pointers, CCFI stores a
cryptographically-secure authentication code in the pointer's unused
most significant bits.  Checking the authentication code of a pointer
before any indirect branches prevents control-flow hijacking because
the attacker cannot compute a valid authentication code without
access to keys. As we will show in Section~\ref{sec:evaluation}, to achieve low overheads
with this scheme, it is essential to have $64$-bit architecture and
to apply the solution to only a subset of the pointers: full-application
of the idea on a~$32$-bit processor results in~$91\%$ overhead for SPEC2017.  In contrast, we
want to enable security for 16, 32- and 64-bit systems, as non-$64$-bit
systems are widely used in Internet-of-Things and Cyber Physical
Systems. Thus, there is a need for new low overhead deployable solutions.

\subsection{\pname{} for CRA Protection} \label{sub:cra_pro}
 \pname{} mitigates
ROP by \textit{ensuring that the addresses of the ROP gadgets in
the gadget chain change after the chain is built}. This will result
in undefined behavior of the payload (likely leading to a program
crash). Consider the example in Figure~\ref{fig:IsoVM}:
\pname{} simultaneously populates multiple (apparent) phantoms of the
program code in the phantom name space; 
to successfully thwart the ROP gadget chain, the
location of the ROP gadgets in all phantoms should be
different~\cite{isomeron2015}.

Traditional in-place randomization techniques~\cite{smash2012, Davi2013:GadgeMe} 
can be used to generate~\phanNO{}s. However, using an aggressive 
randomization approach will complicate
the inverse mapping function,~$f^{-1}$, which is responsible for
recovering the archetype basic block from the different~\phanNO{}s.
This will cause performance overheads with almost no additional
security (beyond changing the gadget addresses in the phantom copies).
\pname{} adopts a more efficient code layout randomization technique
by introducing a security shift,~\SecurityShift{}, between the
individual~\phanNO{}s, so that
they are not perfectly overlapped after removing the phantom
offset,~\DomainShift{}. This simplifies~$f^{-1}$ computations (as shown
in Figure~\ref{fig:PASmask} and maintains code locality.

While the program is executing, \pname{} randomly decides which
copy of the program should be executed next. Figure~\ref{fig:Iso}(a)
shows the normal execution of a program, where~\texttt{Inst 10}
changes the control-flow of the program to a different BBL (starting
with~\texttt{Inst 71}).  After the called BBL is executed, the
control-flow is transmitted to the original landing point~(\texttt{Inst
11} via a \texttt{ret} instruction).  Figure~\ref{fig:Iso}(b) shows
a successful CRA via ROP, in which the attacker uses
a memory safety vulnerability to overwrite the return address stored
on the stack and divert the control flow to~\texttt{Inst 24} upon
executing the~\texttt{ret} instruction. Figure~\ref{fig:Iso}(c)
shows the diversified execution of a program with~\pname{}. For
simplicity, we only show two phantoms and use a security 
shift,~\SecurityShift{}, sized to one instruction. 
Each control flow instruction can arbitrary choose
to change the execution domain or not. Here, the~\texttt{Randomize}
operation decides to execute \texttt{Inst 71} from the~\phanNO{} domain.
As the attacker cannot predict this runtime decision in advance,
\ze{} provide the wrong gadget address on the stack (now shifted
by~\SecurityShift{}). Thus, \ze{} will end-up executing a
\texttt{WRONG} instruction, as shown in Figure~\ref{fig:Iso}(d).
This~\texttt{WRONG} instruction may belong to a different BBL or
divert the execution to a new undesired BBL. In general, if the
attacker makes the wrong guess, they will execute one less (or one
more) instruction compared to the desired gadget. If~\SecurityShift{}
is smaller than the instruction size, the attacker will skip a
portion of the instruction resulting in an incorrect
instruction decoding.

\section{\pname{} Security Extensions} \label{sec:enhancements}

In this section, we discuss two~\pname{} enhancements that boost
its security guarantees against state-of-the-art CRAs.

\subsection{\TRAP{} Instructions} \label{sub:trapinstructions}

\begin{figure}[!t]
  \centering
  \includegraphics[width=0.38\textwidth]{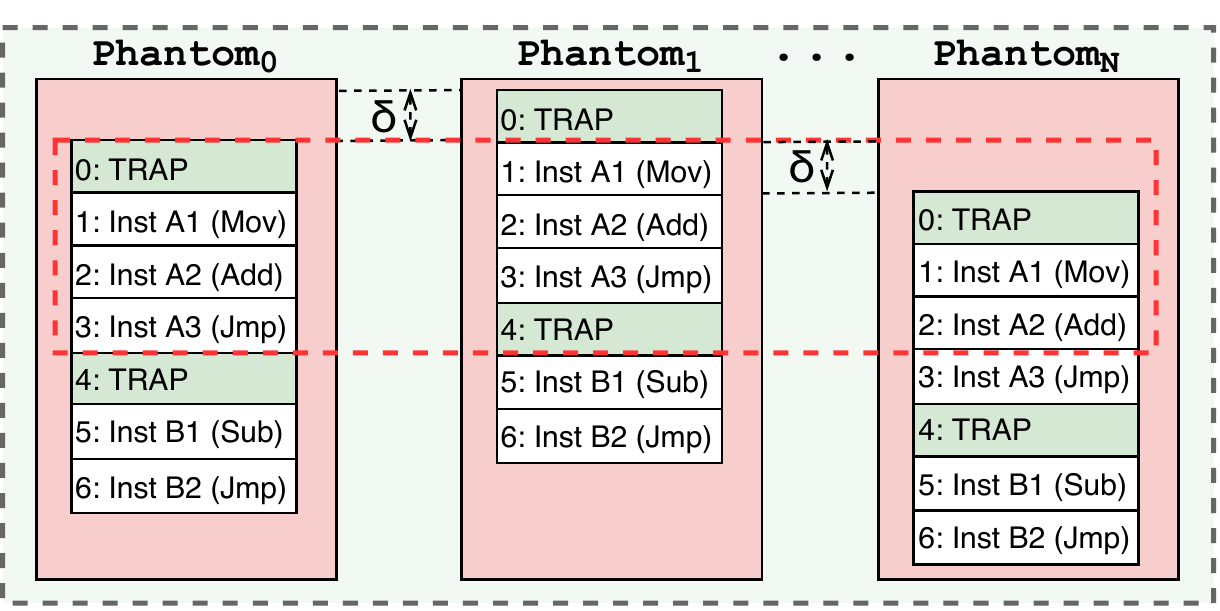}
  \caption{\pname{} with~\TRAP{} instructions.}\label{fig:PAStrap}
\end{figure}

To further limit the attack surface of \pname{}, we add \TRAP{}
instructions. These instructions are inserted at the beginning of
every \BBLEx{}. While \pname{} is enabled, the security
shift,~\SecurityShift{}, will cause the \TRAP{} instruction that
exists in the beginning of a \BBL{} in~\origNO{} domain to appear
at different locations of the same \BBL{} in each of the~\phanNO{} domains, as shown in
Figure~\ref{fig:PAStrap}. This way we provide the ability to catch
attackers that guess the incorrect diversification on the \TRAP{}
instruction while targeting~\BBL{} boundaries.

Let us consider the example of Figure~\ref{fig:Iso}(d). Here, the 
attacker tries to divert the control flow to~\texttt{Inst 24} upon executing 
the~\texttt{ret} instruction. 
Our~\texttt{Randomize} operation decides to execute~\texttt{Inst 71} from 
the~\phanNO{} domain. As a result, the attacker will step on the instruction 
that follows~\texttt{Inst 24} in~\phanNO{}, which will be a \TRAP{} instruction 
in the new model. Executing a \TRAP{} instruction results in a security exception and program 
termination, effectively thwarting the attacker. Programs never execute \TRAP{} 
instructions in normal conditions (no attack) as there exist no 
control-flow transfers to them.\footnote{We modify the hardware to handle 
the case of a fall-through BBL to prevent 
legitimately stepping on a \TRAP{} instruction.}

\subsection{\PTRENCNormal{}} \label{sub:ptrenc}

Besides ROP, CRA variants also extensively rely on pointer corruption
(e.g, JOP/COOP~\cite{Bletsch2011:JOP, Schuster2015})
to subvert a program's intended control flow. There also exist many 
software-based mitigations for JOP/COOP-like attacks~\cite{llvm-cfi,
Chao2016, Bounov2016, Burow2018}. In this paper, we use a
hardware-based technique for hardening \pname{} against them. Since
the attacker needs to overwrite legitimate pointers used by indirect
branches to launch the attack, we encrypt the contents of the pointer
upon creation and only decrypt it upon usage (at a call site).
Consequently, attackers cannot correctly overwrite it.

To achieve the above goal our \PTRENCNormal{} (\PTRENCShort{}) scheme
adds two new instructions:
\ENCP{} and \DECP{}. The two instructions can either be emitted by the compiler
(if re-compiling the program is possible) or inserted by a binary rewriter.

\noindent \textbf{Encrypt Pointer} (\ENCPex{})\textbf{.} The mnemonic \ENCP{}
indicates an encryption instruction. \texttt{RegX} is the register containing
the pointer, e.g., virtual function pointers. The register that holds the encryption
key is hardware-based and never appears in the program binary.

\noindent \textbf{Decrypt Pointer} (\DECPex{})\textbf{.} The mnemonic \DECP{}
indicates a decryption instruction. \texttt{RegX} is the register containing the
pointer. The
register that holds the decryption key is hardware-based and does not appear in
the program binary. As a result, the attacker cannot directly leak the key's
value. Moreover, the attacker cannot simply use the new instructions as signing 
gadgets to encrypt/decrypt arbitrary pointers as \ze{} will have to hijack the 
control flow of the program first.

Unlike prior pointer encryption solutions, which use weak XOR-based 
encryption~\cite{Cowan2003:PointGuard, Tuck2004:PointerEncryption},~\pname{} 
relies on strong cryptography (The QARMA Block Cipher 
Family~\cite{Avanzi2017:QARMA}). 
In contrast to full CCFI solutions~\cite{Mashtizadeh2015, Qualcomm2017},
which use pointer authentication to protect
all code pointers including return addresses, our approach only
guards pointer usages (loads and stores).  Return addresses are
handled by \pname{} randomization, reducing the overall performance
overheads, as will be shown in Section~\ref{sec:evaluation}.

\section{Performance Evaluation}\label{sec:evaluation}

\begin{figure*}[!th] 
  \centering
  \includegraphics[width=0.9\textwidth]{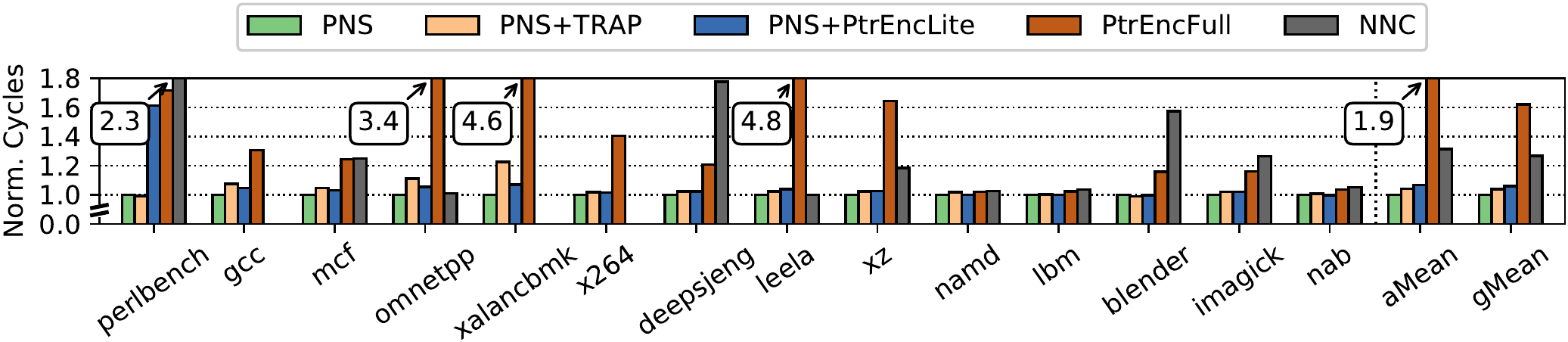}
  \caption{\pname{} performance evaluation for SPEC2017 C/C++ benchmarks.\textsuperscript{\ref{foot:swiso}}}\label{fig:eval}
\end{figure*} 

In this section, we compare \pname{} against prior
solutions, and devise experiments to
analyze the benefits of each aspect of the \pname{} scheme.

As we focus on resource constrained (wimpy) devices, we use \texttt{ARM}
ISA to demonstrate \pname{} as it dominates the embedded and mobile
markets with its 32-bit \texttt{ARMv5-8} instruction set architecture
(ISA).  However, the concept of \pname{} can be applied to any other
ISA (e.g., RISC-V).

\begin{table}[t] %
  \centering
  \scalebox{1}{\begingroup
  \begin{tabular}{r l}
  \toprule
  Core                & ARMv7a OoO core at 1.8 GHz \\
                          & BPred:  BiModeBP, 4096-entry BTB, 48-entry RAS \\
                          & Fetch:      3 wide, 48-entry IQ\\
                          & Issue:      8 wide, 60-entry ROB \\
                          & Writeback:  8 wide, 16-entry LQ, 16-entry SQ \\
  \midrule
  L1 I-cache      & 32KB, 2-way, 2 cycles, 64B blocks, LRU replacement, \\
                          & 2 MSHRs, no prefetch                   \\
  \midrule
  L1 D-cache      & 32KB, 2-way, 2 cycles, 64B blocks, LRU replacement, \\
                            &  16-entry write buffer, 6 MSHRs, no prefetch                 \\
  \midrule
  L2 cache            & 2MB, 16-way, 15 cycles, 64B blocks, LRU replacement, \\
                             &  8-entry write buffer, 16 MSHRs, stride prefetch                 \\
  \midrule
  DRAM                & LPDDR3, 1600 MHz, 1GB, 15ns CAS latency and row \\
                            & precharge, 42ns RAS latency                         \\
  \bottomrule
  \end{tabular}
  \endgroup
  }
    \caption{Simulation parameters.}\label{tab:config}
\end{table} %

\subsection{Experimental Setup} \label{subsec:setup}

We implement \pname{} in the out-of-order (OoO) CPU model of
Gem5~\cite{gem5} for the \texttt{ARM} architecture.  We execute
\texttt{ARM32} binaries from the SPEC CPU2017~\cite{Bucek2018:SPEC}
C/C++ benchmark suite on the modified simulator in syscall
emulation mode with the \texttt{ex5\_big} configuration (see Table~\ref{tab:config}), which is
based on the \texttt{ARM} Cortex-A15 32-bit processor.

To compile the benchmarks, we build a complete toolchain based on a modified 
Clang/LLVM v7.0.0 compiler including \texttt{musl}~\cite{musl}, 
\texttt{compiler-rt}, \texttt{libunwind}, \texttt{libcxxabi}, and 
\texttt{libcxx}. Using a full toolchain allows us to instrument all binary code 
including shared libraries and remove them from the trusted code base (TCB). In 
order to evaluate~\pname{}, we use our modified toolchain to generate the 
following variants.

\fakesubsub{\ToolchainBaseline{}.} This is the case of an unmodified 
unprotected machine. Specifically, we compile and run the SPEC CPU2017 benchmarks 
using an unmodified version of the toolchain and Gem5 simulator. In all of our 
experiments, we use the total number of cycles (numCycles) to complete the 
program, as reported by Gem5, to report performance. The numCycles values of the 
defenses are normalized to this baseline implementation without defenses; thus, 
a normalized value greater than one indicates higher performance overheads.

\fakesubsub{\ToolchainPasNoTrap{}.} In this scenario, we run unmodified binaries on 
our modified Gem5 implementation with all optimizations, as described in 
Section~\ref{sec:microarchitecture}. 

\fakesubsub{\ToolchainPasTrap{}.} In order to prepare \ToolchainPasTrap{}
binaries, we implement an LLVM backend pass to insert \texttt{TRAP}s
at the beginning of BBLs.  This step can also be achieved with an
appropriate binary rewriter making it compatible with legacy
binaries~\cite{Kim2017, Ha2018, williams2020egalito}.
In our compiler pass,
we modified our~\pname{} address translation function in Gem5 to
avoid executing the inserted \texttt{TRAP} instructions for normal
program execution while resolving the~BBL$_{\phanNO{}s}$ 
correctly to a single BBL in the virtual 
address space.

\fakesubsub{\ToolchainPasccfi{}.} To evaluate the performance of~\pname{} 
with \PTRENCShort{}, we first write an LLVM IR pass to instrument the code 
(including shared libraries) and insert the relevant instructions as described 
in CCFI~\cite{Mashtizadeh2015}. Specifically, we emit instructions whenever
(1)~a new 
object is created (to encrypt the contents of the \texttt{vptr}), (2)~a virtual 
function call is made (to decrypt the \texttt{vptr}), or (3)~any operation on 
code pointers in C programs. Then, we appropriate the encodings for 
\texttt{ARM}'s {\tt\small ldc} and {\tt\small stc} instructions respectively, 
which are themselves unimplemented in Gem5, to behave as~\ENCP{} and~\DECP{} 
instructions. We add a dedicated functional unit in Gem5 to handle these 
instruction's latency in order to avoid any contention on the regular functional 
units. We also assume equal cycle counts of~$8$ for both 
instructions~\cite{Avanzi2017:QARMA}. This 
latency is to emulate the effect of the actual encryption/decryption.

\fakesubsub{\ToolchainCcfifull{}.} In this approach, we instrument code 
pointer load/store operations in addition to function entry/exit points to 
protect return addresses for non-leaf functions. Conceptually, this solution is 
similar to \texttt{ARM} PAC~\cite{Hans2018}. However, 
due to the absence of PAC support in Gem5 (and for~$32$-bit \texttt{ARM} 
architectures in general), we only perform behavioral simulation for comparison 
purposes, without keeping track of the actual pointer metadata.

\fakesubsub{\ToolchainSwisoEx~(\ToolchainSwiso{}).} For the sake of completeness and
fair comparison, we also implement a static version where there are
two copies of the code, i.e., a version without the phantom aspect
of the naming scheme.  In this model, we have two virtual addresses for
each instruction but these addresses are physically stored in memory,
essentially halving the capacity of the microarchitectural
structures.\footnote{This model is similar to the Isomeron 
solution proposed by Davi \emph{et al.}~\cite{isomeron2015}, with the
modification that it is used at the \BBL{} granularity as opposed
to the original work, which uses a dynamic binary rewriting framework
with function granularity.}  We create the two copies by introducing
a shift of~\texttt{TRAP} instruction size in one of them.  At a
high-level our implementation works as follows: (1)~clone functions
using an LLVM IR pass, (2)~LLVM backend pass to insert \texttt{TRAP}s
for cloned functions, (3)~instruct the LLVM backend to globalize
\BBL{} labels, (4)~emit a diversifier \BBL{} for every \BBL{}, and
(5) rewrite branch instruction targets to point to the
diversifier.

Of the~$16$ \texttt{C/C++} benchmarks,~$14$ compile with all different toolchain 
modifications. \texttt{parest} has compatibility issues with \texttt{musl} due 
to exception handling usages, while \texttt{povray} failed to run on Gem5. 
For~\ToolchainSwiso{}, \texttt{gcc}, \texttt{xalancbmk}, and \texttt{x264} 
present compilation and/or linking issues.

\subsection{Experimental Results}

We run all benchmarks to completion with the \texttt{test} input set on our 
augmented Gem5. We verified the correctness of the outputs against the reference
output. Figure~\ref{fig:eval} shows the performance overhead of the different 
design approaches (all normalized to~\ToolchainBaseline{}). As 
expected,~\ToolchainPasNoTrap{} has identical performance 
to~\ToolchainBaseline{}. The overhead of~\ToolchainPasTrap{} is 
minimal,~$0\%$--~$22\%$ (avg.~$4\%$)\footnote{This average 
is for~$14/16$ benchmarks, as explained in Section~\ref{subsec:setup}}. 
Adding support for~\PTRENCShort{} increases the 
performance overheads of~\ToolchainPasccfi{} 
to~$0\%$--~$61\%$ (avg.~$6\%$). 
The \texttt{perlbench} benchmark suffers from a relatively high 
overhead due to its extensive use of function pointers and indirect branches. On the 
other hand, fully protecting the 
binaries with a deterministic defense such as~\ToolchainCcfifull{} encounters 
a~$91\%$ overhead on average (geometric mean of~$62\%$). Our static 
implementation of software \ToolchainSwiso{} introduces an arithmetic average overhead 
of~$31\%$ (geometric mean of~$26\%$)\footnote{\label{foot:swiso}Our 
\ToolchainSwiso{} implementation does not instrument external 
libraries (only the main application code) due to compilation issues. 
This leads to overheads that are less than 
intuitively expected.}. In contrast to software Isomeron~\cite{isomeron2015} which relies on dynamic
binary instrumentation (DBI), the overheads 
for our implementation are primarily attributed to the indirection every BBL 
branch must make to the diversifier.

\begin{figure}[!t] 
  \centering
  \includegraphics[width=0.45\textwidth]{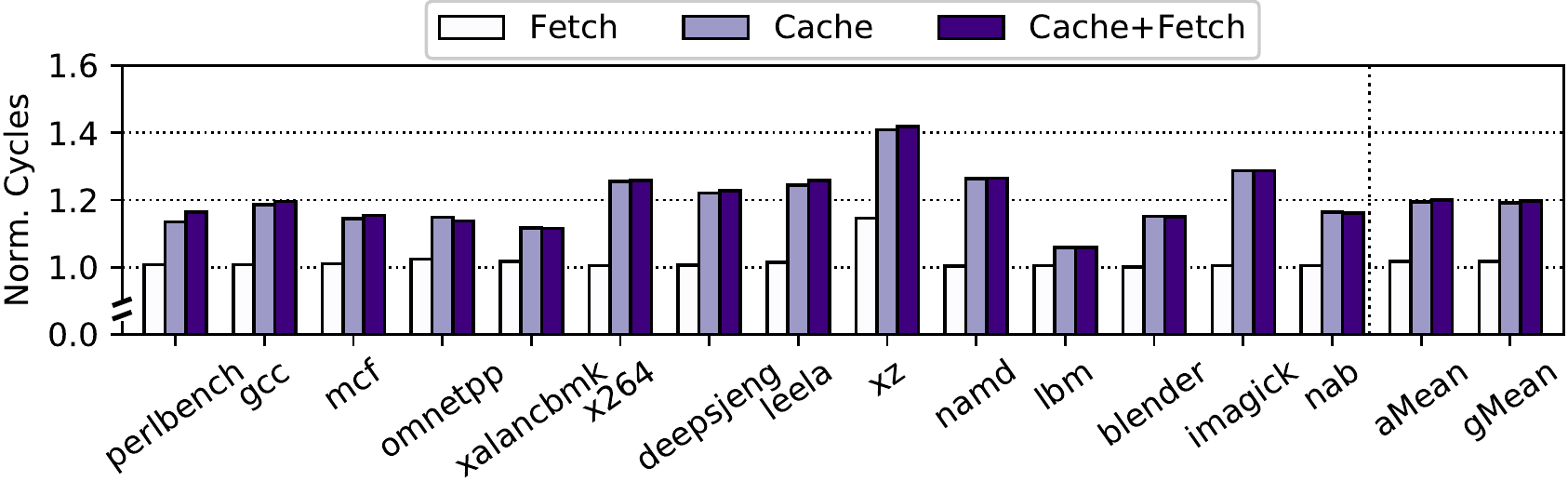}
  \caption{\pname{} performance evaluation with additional one-cycle access latency for fetch stage, L1-I\$, and both.}\label{fig:eval-oneCycle}
\end{figure}

As illustrated in Section~\ref{sec:microarchitecture}, the 
required~\ToolchainPasNoTrap{} modifications do not add additional cycle latency 
to the processor pipeline. However, we performed an additional set of experiments with 
a more conservative assumption of having one additional cycle latency for all 
instructions in fetch stage, or one more cycle for accessing L1 instruction 
cache, or both. We show results compared to an unmodified baseline in 
Figure~\ref{fig:eval-oneCycle}. We notice an average performance overhead 
of~$1\%$ for stalling the fetch stage. However, stalling the instruction cache for 
one cycle (hit latency is originally two cycles) is more harmful to the 
performance. Thus, the I\$ optimizations are mandatory, as described in 
Section~\ref{sec:microarchitecture}.

Finally, the call depths listed in Table~\ref{tab:call-depth} show that SPEC 
programs do not exceed a depth of~$244$ (\texttt{leela}), indicating that 
a~$256$-entry hardware~\ShadowStackEx{} is sufficient to handle the common execution 
cases.

\begin{table}[h!] 
  \centering
  \scalebox{1}{\begingroup
    \begin{tabular}{@{} l l | l l | l l l@{}}
    \toprule
    Bench. & Call & Bench. & Call & Bench. & Call   \\
    Name & Depth & Name & Depth & Name & Depth   \\ \midrule
    perlbench      & 24         & x264           & 15         &    lbm            & 10      \\
    gcc            & 28         & deepsjeng      & 48         &    blender        & 23       \\
    mcf            & 28         & leela          & 244        &      imagick        & 22      \\
    omnetpp        & 196        & xz             & 16         &    nab            & 16       \\
    xalancbmk      & 77         & namd           & 12         &   povray         & -          \\ 
    \bottomrule
    \end{tabular}
  \endgroup
  }
    \caption{Max. call depth for SPEC C/C++ benchmarks.}\label{tab:call-depth}
\end{table}

\subsection{FPGA Prototyping}

For the sake of completeness, we have developed an FPGA prototype of~\pname{}
using the Bluespec hardware description language (HDL). Specifically, we
added~\pname{} hardware modifications to the front-end of the~$32$-bit
\texttt{Flute} RISC-V processor, a 5-stage in-order pipelined processor
typically used for low-end applications that need MMUs~\cite{Flute}.
We prototyped the processor on the the Xilinx Zynq (ZCU106) Evaluation Kit.

Our evaluation results shows that we can reliably run with a clock period
of~$7.5$ ns (maximum frequency of~$133$ MHz) for both the baseline core and the
modified one. The area increase due to~\pname{} is negligible~($0.83\%$ extra
Flip-Flops with~$2.02\%$ additional LUTs). We verified the correctness of our
FPGA implementation by running simple bare-metal applications. Due to the lack
of mature support for~$32$-bit OSs on RISC-V, we were not able to run the
application benchmarks we used in simulation. We leaving booting Linux on
our modified core as future work.

\section{Security Analysis \& Evaluation}\label{sec:securityanalysis}
In this section, we present our evaluation results regarding the effectiveness
of \pname{} and discuss its security guarantees against CRAs.  

\subsection{Threat Model}
\noindent\textbf{Adversarial Capabilities.} We consider an adversary
model that is consistent with previous work on code-reuse attacks
and mitigations~\cite{Mashtizadeh2015, isomeron2015, 
Gallagher2019:Morpheus, Schuster2015}.  We assume that
the adversary is aware of the applied defenses and has access to
the source code, or binary image, of the target program.  Furthermore,
the target program suffers from memory safety-related vulnerabilities
that allow the adversary to read from, and write to, arbitrary
memory addresses. The attacker's objective is to (ab)use memory
corruption and disclosure bugs, mount a code-reuse attack, and achieve privilege
escalation.

\noindent\textbf{Hardening Assumptions.} We assume that the underlying OS is 
trusted. If the OS is compromised and the attacker has kernel privileges, the attacker can execute malicious code
without making ROP-style attacks; a simple 
mapping of the data page as executable will suffice. We assume that ASLR 
and W\textasciicircum{X} protection are enabled---i.e., no code injection is allowed (non-executable
data), and all code sections are non-writable (immutable code).
Thus, attacks that modify program code at runtime, such as
rowhammer~\cite{Kim2014}, are out of scope.  We also do not consider
non-control data attacks~\cite{Szekeres2013}, such as Data-Oriented
Programming~\cite{Hu2016} and Block-Oriented 
Programming~\cite{Ispoglou2018:BOP}. This class of attacks only tamper-with 
memory load and store operations, without inducing any unintended
control flows in the program.  This limitation also applies to prior 
work as well~\cite{Mashtizadeh2015, isomeron2015, burow2017control, Gallagher2019:Morpheus}.
Lastly, every other standard hardening feature (e.g.,~stack-smashing
protection~\cite{cowan1998stackguard}, CFI~\cite{burow2017control}) is \emph{orthogonal} to \pname; our proposed
scheme does not require nor preclude any such feature.

\subsection{Secrets}

There are no secret parameters in the basic \pname{} scheme. The
number of phantoms and the security shift can be made public
as security comes from the random selection of names. For
\pname{} extensions, a per-process key (used for encryption)
should be kept secret for the lifetime of the respective process.

\subsection{Quantitative Security Analysis}

\fakesubsub{ROP-Gadget Chain Evaluation.} To evaluate~\pname{} 
against real-world ROP attacks we use Ropper~\cite{Ropper}, a tool that can 
find gadgets and build ROP chains for a given binary. A common ROP attack is to 
target the \texttt{execve} function with \texttt{/bin/sh} as an input to launch a 
shell. As the chain-creation functionality in Ropper is only available for 
\texttt{x86}~\cite{Ropper}, we analyze SPEC2017 \texttt{x86} 
binaries for this particular exploit and report the number of available gadget 
chains ($\overline{\pname{}}$). 

To emulate the effect of~\pname{}, we 
modified the Ropper code to extend each gadget length by one byte, decode the 
gadget, and check if the new gadget is semantically equivalent to the old one or 
not. This emulates the effect of an attacker targeting a particular address, but 
instead executing the one before due to the~\pname{} security 
shift,~\SecurityShift{}. As shown in Table~\ref{tab:pas-rop},~\pname{} foils 
all the gadget-chains found by our modified Ropper. Extending the Ropper chain-creation 
functionality to the ARM ISA is part of our future work. Intuitively, the results 
would be even worse for the attacker in ARM as the state-space is more 
constrained due to instruction alignment requirements. 

\begin{table}[t]
 	\centering
  	\scalebox{0.8}{\begingroup
	\begin{tabular}{@{} l l l | l l l | l l l@{}}
              \toprule
              Bench. & $\overline{\pname{}}$ & $\pname{}$ & Bench. & $\overline{\pname{}}$ & $\pname{}$ & Bench. & $\overline{\pname{}}$ & $\pname{}$ \\ 
              Name      & Chains             & Chains  & Name      & Chains             &  Chains   & Name      & Chains             &  Chains \\                                
              \midrule
              perlbench & 17                 & 0       & x264      & 23                 & 0       & lbm       & 23                 & 0       \\
              gcc       & 23                 & 0       & deepsjeng & 11                 & 0       &blender   & 23                 & 0      \\
              mcf       & 11                 & 0       & leela     & 15                 & 0 & imagick   & 23                 & 0       \\
              omnetpp   & 23                 & 0       & xz        & 11                 & 0   & nab       & 23                 & 0      \\
              xalancbmk & 15                 & 0       & namd      & 23                 & 0  & povray    &  23                & 0      \\
              \bottomrule
    \end{tabular}    
    \endgroup
    }
     \caption{ROP gadget-chain reduction for SPEC2017 C/C++ benchmarks. $\overline{\pname{}}$ and $\pname{}$ correspond to the number of valid ROP chains before and after~\pname{}.}\label{tab:pas-rop}
\end{table}

\fakesubsub{Control-flow Hijacking Evaluation.} We further evaluate 
security by using RIPE~\cite{Wilander2011:RIPE}, an open source
intrusion prevention benchmark suite. We port RIPE to ARM and run
it on our modified Gem5, with~$n = 8$ bits, as described in Section~\ref{sec:evaluation}. 
We mainly focus on return-address manipulation as a target code
pointer and \texttt{ret2libc}/ROP as attack payloads. Shellcode
attacks are not considered as we 
expect W\textasciicircum{X}.

Our ported RIPE benchmark contains~$54$ (relevant) attack combinations. 
On an unprotected Gem5 system,~$50$ attacks succeed and 4 attacks
fail. After deploying~\pname{}, all of the~$54$ attacks fail including the 
single-gadget \texttt{ret2libc} attacks. That is mainly due to our high number 
of phantoms present at runtime,~$2^8=256$.

That said, real-world exploits typically involve payloads with
several gadgets. According to Cheng \emph{et al.}~\cite{Cheng2014} the
shortest gadget chain consists of thirteen gadgets. Hence, the
probability for successful execution of a gadget chain is~$p_{success}
\leq \left(\frac{1}{256}\right)^{13} = 4.93 \times 10^{-32}$. 
Snow \emph{et al.}~\cite{snow2013} successfully
exploited a vulnerability with a ROP payload consisting of only six
gadgets, which would equate to a better, but still low, success
probability of~$p_{success} = \left(\frac{1}{256}\right)^{6} = 3.55 \times 10^{-15}$.

\subsection{Qualitative Security Evaluation}

\fakesubsub{Just-In-Time Return-Oriented Programming.}
Although JIT-ROP~\cite{snow2013} permits the attacker to construct a compatible
code-reuse payload \textit{on the fly}, \ze{} cannot modify the gadget chain
after the control flow has been hijacked. As a result, the attacker needs to
guess the domain of execution of the entire JIT-ROP gadget-chain in advance.
So,~\pname{} mitigates JIT-ROP similarly to how it mitigates (static)
ROP/JOP/COP: i.e., by removing the attacker's ability to put together (either
in advance or on the fly) a valid code-reuse payload. The above security guarantees 
are achieved by the regular~\pname{} proposal (as explained in 
Section~\ref{subsec:pas-cra}) with no extensions or program recompilation, 
making it suitable for legacy binaries and shared third party libraries.

\fakesubsub{Blind Return-Oriented Programming.} BROP attacks can remotely find 
ROP gadgets, in network-facing applications, without prior knowledge of the 
target binary~\cite{Bittau2014:HackingBlind}. The idea is to find enough gadgets 
to invoke the \texttt{write} system call through trial and error; then, 
the target binary can be copied from memory to the network to find more gadgets. 
As a proof of concept, the authors showed an example with~$5$-gadgets that 
invokes \texttt{write}. With~\pname{}, the success probability of invoking 
\texttt{write} would be~$\left(\frac{1}{256}\right)^{5} = 9.09 \times 10^{-13}$. 
Note that completing an end-to-end attack requires harvesting, and using, even 
more gadgets, after dumping the target binary, which makes the attack unfeasible 
on a PNS-hardened system. Additionally, BROP requires 
services that restart after a crash, while failed attempts will be noticeable to 
a system admin. 

\fakesubsub{Whole-function Reuse.}
Unlike ROP attacks, which (re)use short instruction sequences, entire
functions are invoked, in this case, to manipulate the control-flow of the
program. This type of attack includes counterfeit object-oriented programming
(COOP) attacks, in which whole C++ functions are invoked through code pointers
in read-only memory, such as \texttt{vtable}s~\cite{Schuster2015}. \pname{}
relies on the \PTRENCShort{} extension to prevent the attacker from manipulating pointers
(\texttt{vptr}) that point to \texttt{vtable}s---a necessary step for mounting
a COOP attack. 

\texttt{Ret2libc} is another example for whole function reuse attacks, in
which the attacker tries to execute entire \texttt{libc}
functions~\cite{ret2libc_solar:1997, Nergal2011}.\footnote{In general, 
any function, of
any other shared library, or even the main binary itself, can be used
instead.} With~\pname{}, the attacker will have to guess the address of the 
first basic block of the function in order to lunch the attack, reducing the 
success probability to~$\left(\frac{1}{256}\right) = 0.0039$. 

Our analysis of real-world exploits shows that executing a \texttt{ret2libc}
attack incurs multiple steps in order for the attacker to (1)~prepare the
function arguments based on the calling convention, (2)~jump to the desired
function entry, (3)~silence any side-effects that occur due to executing the
whole function, and (4)~reliably continue (or gracefully terminate) the victim
program without noticeable crashes. (1) and (3) generally requires code-reuse
(ROP) gadgets, as demonstrated by the following publicly-available exploits:
(a)~ROP + \texttt{ret2libc}-based exploit against
mcrypt~\cite{exploitDB:mcrypt}, (b)~ROP +
\texttt{ret2libc}-based exploit against
Nginx~\cite{exploitDB:Nginx}, (c) ROP +
\texttt{ret2libc} + shellcode-based exploit for Apache +
PHP~\cite{exploitDB:Apache} and (d) ROP +
\texttt{ret2libc}-based exploit against
Netperf~\cite{exploitDB:Netperf}. Thus, if the ROP 
part of the exploit requires~$G$ gadgets, the probability for successfully 
exploiting the program would exponentially decrease 
to~$p_{success} \leq \left(\frac{1}{256}\right)^{G}$. That is because the 
attacker will have to guess the domain of execution (out of~$2^8 = 256$ phantoms) of 
every gadget.

A natural question to ask is: can the attacker leverage~\pname{}'s mechanisms to 
hijack the system? The answer is no for the following reasons. 
To divert the control flow of a program, an attacker must corrupt either (1) 
return addresses, which are protected with~\pname{} randomization, or (2) 
function pointers, which are protected by the \PTRENCShort{} extension. To corrupt return 
addresses, an attacker must make a guess (this will exponentially scale with the 
number of return addresses to be corrupted) to determine the correct execution 
domain. To bypass \PTRENCShort{}, an attacker has to leak the key, which is 
hardware-based, or divert the control flow to a signing gadget (an encryption 
instruction). The latter requires hijacking the control-flow first, which is 
already guarded with randomization and encryption thus constructing a chicken 
and egg dilemma.

\fakesubsub{Side-channel Attacks.}
\pname{} takes multiple steps to be resilient to side channel attacks.
Firstly, \pname{} purposefully avoids timing variances introduced due to
hardware modifications, in order to limit timing-based side channel attacks.
Additionally, the attacker cannot leak the random phantom index,~$p$, 
which are generated by the selector as it is
unreadable from both user and kernel mode---it exists within the processor
only. Similarly, the execution domain cannot be
leaked to the attacker through the architectural stack, as \pname{} keeps it
within the hardware in the~\ShadowStackShort{}.

\section{\pname{} System Level Support}\label{sec:deployment}

For completeness, we outline design changes required
to deploy a \pname{} general-purpose system.

\fakesubsub{Sizing.} Although~\ShadowStackShort{}
only stores eight bits per return address in hardware, it still has
a limited size that cannot be dynamically increased as the
architectural stack. This means programs with deeply nested function
calls may result in a~\ShadowStackShort{} overflow. To handle this
issue, we add two new hardware exception signals:
\textit{hardware-stack-overflow} and \textit{hardware-stack-underflow}.
The former is raised when the~\ShadowStackShort{} overflows. In this
case, the OS (or another trusted entity), encrypts and copies the contents
of the~\ShadowStackShort{} to the kernel memory. This kernel memory
location will be a stack of stacks and every time a stack is full
it will be appended to the previous full stack. The second
exception will be raised when the~\ShadowStackShort{} is empty to
decrypt and page-in the last saved full-stack from kernel memory.

\fakesubsub{Stack Unwinding.} Since addresses are split across the
architectural (software) stack and the ~\ShadowStackShort{} it is vital
to keep them in sync for correct operation. Earlier, we described
how normal LIFO \texttt{call}/\texttt{ret}s are handled. 
In some cases, however, the stack can be reset arbitrarily 
by \texttt{setjmp}/\texttt{longjmp} or C++ exception
handling. To ensure the stack cannot be disclosed/manipulated
maliciously during non-LIFO operations, we change the runtime to
encrypt the ~\texttt{jmp\_buffer} before storing it to memory.
Additionally, we also store the current index of the~\ShadowStackShort{}.
When a \texttt{longjmp} is executed, we decrypt the contents of
the~\texttt{jmp\_buffer} and use the decrypted~\ShadowStackShort{}
index to re-synchronize it with the architectural stack. The same
approach can be applied to the C++ exception handling mechanism by
instrumenting the appropriate APIs.

\fakesubsub{Context Switches.} The~\ShadowStackShort{} of the current
process is stored in the Process Control Block before a context
switch.  In terms of cost, the typical size of the~\ShadowStackShort{}
is~$256$-bytes~($256$ entries, each has 8-bits). Moving this number of bytes between
the~\ShadowStackShort{} and memory during context switch requires
just a few \texttt{load} and \texttt{store} instructions,
which consume a few cycles. This overhead is negligible with
respect to the overhead of the rest of the context switch (which
happens infrequently; every tens of milliseconds).

\fakesubsub{Multithreading.} To support multithreading, the
~\ShadowStackShort{} has to be extended with
a multithreading context identifier, which increases
the size of stack linearly with number of thread contexts that can
be supported per hardware core.

\fakesubsub{Dynamic Linking.} Dynamically-linked shared libraries
are essential to modern software as they reduce program size and
improve locality.  Although most embedded system software (the
primary target in this work) in MCUs is typically statically-linked,
we note that \pname{} is compatible with shared libraries as it can
be fully realized in hardware.  Thus, it does not differentiate
between BBLs related to the main program and the ones corresponding to
shared libraries.  On the other hand, dynamic linking has been
a challenge for many CFI solutions, as control flow graph
edges that span modules may be unavailable statically. CCFI~\cite{Mashtizadeh2015} suffers
from the same limitation as the dynamically shared library code
needs to be instrumented before execution; otherwise, the respective
pages will be vulnerable to code pointer manipulation attacks.

\section{Related Work}\label{sec:relatedwork}

\begin{table*}[t]
  \centering
  \resizebox{0.95\textwidth}{!}{
  \begin{tabular}{r c l c l l c}
    \toprule
    \multicolumn{1}{c}{\textbf{Proposal} } & \multicolumn{1}{c}{\textbf{Hardware}} & \multicolumn{1}{c}{\textbf{Software}}      & \multicolumn{1}{c}{\textbf{Randomization}} & \multicolumn{1}{c}{\textbf{Main Sources of Overheads}}  & \multicolumn{1}{c}{\textbf{Cost of Portability to~$32$-bit systems}}  & \multicolumn{1}{c}{\textbf{Energy}}  \\
                                                            & \multicolumn{1}{c}{\textbf{Support}}  & \multicolumn{1}{c}{\textbf{Modifications}} & \multicolumn{1}{c}{\textbf{Interval}}      &                                                                           &                        & \multicolumn{1}{c}{\textbf{Overheads}} \\
    \midrule    
	CaRE~\cite{Nyman2017}                    & Yes          & Recompile    & No                  & Directing every branch to external monitor                & None                                    & Moderate \\
	Intel CET~\cite{IntelCeT}                & Yes          & Recompile    & No                  & Maintaining full shadow stack                             & None                                    & Low \\
	\midrule   
	CCFI~\cite{Mashtizadeh2015}              & No           & Recompile    & No                  & Using complete pointer authentication                     & Extra Load/Store per pointer            & Moderate \\
	ARM PAC~\cite{Qualcomm2017}              & Yes          & Recompile    & No                  & Using complete pointer authentication (negligible on h/w) & Extra Load/Store per pointer            & Moderate \\
	\midrule   
	NVX~\cite{Berger2006:DieHard ,Lu2018:Buddy}   & No           & Recompile          & No   & Running~$N$ program copies simultaneously  & Increase overheads by a factor of~$N$                  & High \\ 
    Isomeron~\cite{isomeron2015}             & No           & DBI          & 1 ms (Func. time)   & Maintaining two program copies (high TLB and I\$ misses)  & None                                    & High \\ 
	Shuffler~\cite{David2016:shuffler}       & No           & DBI          & 50 ms               & Offloading computations to another core/thread            & Double overheads on single-core systems & High \\
    Morpheus~\cite{Gallagher2019:Morpheus}   & Yes          & Recompile    & 50 ms               & Adding 2-bit tags per 64-bit words (pointer size)         & Double memory tags overhead             & Low \\

	\midrule
	\textbf{\pname{}}                        & Yes          & None         & 10 ns (BBL time)    & Using \PTRENCNormal{}                                     & None                                    & Low \\
	\bottomrule
  \end{tabular}
    }
      \caption{Comparison with prior work.}\label{tab:comparison}
\end{table*}

As shown in Section~\ref{sec:introduction}, the idea of having multiple names
for the same instruction is fundamentally different compared to other security 
paradigms. Further, in Section~\ref{sec:evaluation} 
we showed that \pname{} has lower overheads compared with
the state-of-the-art commercial solution (ARM PAC).
In this section, we explore prior CRA mitigations and discuss their benefits and differences
(summarized in Table~\ref{tab:comparison}).

\fakesubsub{CRA Mitigations for Resource Constrained Devices.} Nyman
\emph{et al.}~\cite{Nyman2017} introduced \texttt{CaRE}, an
interrupt-aware control-flow integrity (CFI) scheme for low-end
microcontrollers that leverages TrustZone-M security extensions.
\texttt{CaRE} instruments binaries in a manner which removes all
function calls and indirect branches and replaces them with dispatch
instructions that trap control flow to a branch monitor.  Although
the branch monitor eliminates control-flow attacks that utilize
such branches, the performance overheads introduced range between~$13\%$
and~$513\%$.  In the case of indirect calls, \texttt{CaRE} matches
the branch target against a record of valid subroutine entry points.
Unlike~\ToolchainPasccfi{}, this coarse-grained approach does not
protect against whole function reuse attacks.

\fakesubsub{Hardware-based CRA Mitigations.} Intel architectures
offer a hardware-based CFI technology named Control-flow Enforcement
Technology (CET) that is to be available in future \texttt{x86}
processors~\cite{IntelCeT, Shanbhogue2019:IntelCeTPaper}. CET adds
a new \texttt{ENDBRANCH} instruction, which is placed at the entry
of each BBL that can be invoked via an indirect branch. When an
indirect forward branch occurs, the following instruction is expected
to be an \texttt{ENDBRANCH}, otherwise an attack is assumed. CET
provides only coarse-grained protection where any of the possible
indirect targets are allowed at every indirect control-flow transfer.
Thus, an attacker can still reuse the whole BBL and store the address
of the \texttt{ENDBRANCH} of the desired BBL in the stack as before. 
The above attack will fail against~\pname{} with high probability as every 
instruction (and basic block) can have up to~$N$ different addresses 
forcing the attacker to gamble on which one to use. 

Additionally, CET protects \texttt{call-return} instructions using
a full shadow stack (i.e.,~$32$ or~$64$ bits per entry), that resides in virtual 
memory. Unlike a shadow stack which compares return address 
on every \texttt{ret} instruction, our~\ShadowStackShort{} only concatenates the 
domain bits to the return address with no wasteful comparisons. 
Furthermore,~\pname{} uses a smaller hardware structure (the
~\ShadowStackShort{}) that consumes~$8$ bits per entry and that cannot be leaked by an attacker who 
can illegally tamper main memory.

Recently, ARM introduced the Pointer Authentication Code (PAC) feature in 
\texttt{Armv8.3A} as a hardware primitive to mitigate CRAs~\cite{Qualcomm2017}. 
Hans \emph{et al.} showed how to harden ARM PAC against 
reply attacks by using unique tweaks (along with the authentication key) for 
different pointer types~\cite{Hans2018}. As discussed in Section~\ref{subsec:current-mitigations}, ARM PAC relies on the 
currently unused upper bits of the~$64$-bit pointers. Mapping the same technique 
to non $64$-bit systems results in high performance overheads, as evaluated in 
Section~\ref{sec:evaluation}.

While our~\ToolchainPasccfi{} extension relies on cryptographic algorithms 
similar to ARM PAC~\cite{Qualcomm2017},~\ToolchainPasccfi{} has two main 
advantages. First,~\ToolchainPasccfi{} uses encryption instead of 
authentication to avoid storing additional metadata (authentication code) per 
pointer on~$32$-bit systems. Second, ARM PAC is applied for all code pointers 
including return addresses and function pointers. This is represented 
by~\ToolchainCcfifull{} in our evaluation. On the other 
hand,~\ToolchainPasccfi{} is only applied for function pointers (and C++ virtual 
pointers) as the return addresses are protect by~\pname{}'s fine-grained 
randomization. The reduction in the cryptographically-protected locations highly 
reduced the performance overheads, as shown in Section~\ref{sec:evaluation}. 

\fakesubsub{N-Variant eXecution Systems.} The general idea of 
N-variant execution (NVX) systems
is to run~$N$ \textit{different} copies/variants of the same code, alongside 
each other, while checking their runtime behavior~\cite{Berger2006:DieHard, 
Cox2006}. If the variants produce a different response to a single common input 
(due to an internal failure or external attack payload), the checker detects 
such divergences in execution and raises an alert. Since 2006, many NVX systems 
have been proposed to achieve reliability and security 
goals~\cite{Volckaert2016:DCL, Volckaert2016, Koning2016, Kwon2016:LDX, 
Gawlik2016:Detile, Lu2018:Buddy}. While NVX systems can offer additional benefits 
over~\pname{}, such as precise failure detection, they suffer from 
considerable performance (at least~$100\%$) and memory overheads, and therefore 
are not suitable for resource constrained systems.

\fakesubsub{Code Randomization.} Multiple work has proposed using in-place code 
randomization (aka instruction set randomization) to defeat code reuse 
attacks~\cite{Gaurav2003:ISR,Antonis2013:ASIST,Sinha2017:ISR}. The main idea is 
to randomize the encoding of instructions in memory, while maintaining a unique 
instruction name per program execution. Unlike~\pname{}, which dynamically 
change instruction names at runtime, the above techniques are static (i.e., 
applied in compile time) in nature. As a result, they are susceptible to JIT-ROP 
attacks while~\pname{} is not. 

\fakesubsub{Live Randomization.} Recent work has pioneered the use
of hardware moving target defenses to protect against code-reuse
attacks~\cite{Gallagher2019:Morpheus}.  Gallagher \emph{et al.}
proposed Morpheus, an architecture that (1) displaces code and data
pointers in the address space (2) diversifies the representation
of code and pointers using strong encryption,
and (3) periodically repeats the above steps using a different
displacement and key. 

The main conceptual difference between Morpheus and \pname{} is
that in \pname{}, at any given instant there are multiple names
(addresses) for an instruction while there is only one name (address)
for an instruction in Morpheus. This distinction is also true of
\pname{} and software moving target systems~\cite{David2016:shuffler}
used to protect against code reuse attacks.

In terms of security, Morpheus must keep two parameters a secret
until they are changed: displacements for the code and data regions,
and keys for encrypting/decrypting pointers.  In the basic \pname{}
there are no secrets, and in the enhanced \pname{}, there is only
one secret viz.  the key used for code pointer encryption.  If
\pname{} is added to Morpheus it increases the security level offered
by Morpheus because the attacker has to disclose the displacement
\emph{and} break the name confusion to mount a code-reuse attack.

\pname{} can also provide an illusion of a faster churn rate.  The
churn time can be thought of as the time an attacker has to deploy
a countermeasure.  \pname{}, forces the attacker to have a counter
strategy every basic block which normally completes execution in
the order of nanoseconds.  While Morpheus' churn rate (milliseconds
for \pname{} level of performance) is sufficient to protect against
remote network adversaries, the (apparently) faster churn provided
by \pname{} is meaningful in offering protection against local
attackers especially with side channel capabilities, and thus is
again complementary to Morpheus.  The BBL-by-BBL apparent churn
offered by \pname{} also comes at much lower energy cost compared
to Morpheus as it does not require memory scanning to identify
pointers. Finally from a deployment perspective, a unique benefit of ~\pname{}
is that it works for ``wimpy'' non-64-bit systems while Morpheus and
software moving target systems, rely on the availability of a 64-bit
address space for security.

\fakesubsub{Memory Safety Defenses.} Hardware primitives for memory safety, such 
as LowFat~\cite{Kwon2013:LowFatHW}, CHERI~\cite{Watson2015:CHERI}, 
REST~\cite{Kanad2018:REST}, and Califorms~\cite{Sasaki2019:Califorms}, can 
mitigate CRAs by detecting the initial memory safety violation. While providing 
higher security guarantees than~\pname{}, the above techniques are less suitable 
for resource constrained systems due to their high energy overheads in addition 
to their intrusive changes to the entire software stack (including software, 
hardware, and OS). On the contrary,~\pname{} is able to enhance 
security even for legacy binaries.

\section{Conclusion}\label{sec:conclusion} 

In this paper, we proposed~\pname{}, a name confusion design that
allows for multiple addresses/names for individual instructions.
We also demonstrated an application of \pname{}, which is used to mitigate
code-reuse attacks, a prominent class of attacks that have proven costly to
mitigate. The key idea is to force the attacker to
carry out the difficult task of guessing which randomly-chosen name
will be used, by the hardware, to carry out a successful
attack. Building on this idea we show that we protect
against a variety of code-reuse attacks, including the state-of-the-art JIT-ROP and
COOP.

While it offers strong security guarantees, 
\pname{} requires minor modifications to the processor front-end: 
specifically, it requires changes to indexing functions,
8 metastable flip-flops, and~$256$ bytes of state. Experimental results
showed that~\pname{} incurs negligible performance impact compared
to hardware-based cryptographic control-flow integrity schemes, which
is the state-of-the-art commercial solution (implemented on the
iPhone XS). Another major benefit of~\pname{} is that it
does not depend on ``free'' bits or the vastness of the 64-bit address
space to work, making it suitable for 16- and 32-bit microcontrollers
and microprocessors.
For the foreseeable future, code-reuse attacks will continue to
plague systems security. The increased proliferation of resource-constrained
systems that cannot deal with the performance overheads
of server-grade defenses calls for more efficient mitigations.
Thus, \pname{} provides a cheaply deployable hardware technique
that strengthens control flow protection uniformly across embedded
and server ecosystems.

{
\small
\bibliographystyle{IEEEtranS}
\bibliography{main}

% Generated by IEEEtranS.bst, version: 1.12 (2007/01/11)
\begin{thebibliography}{10}
\providecommand{\url}[1]{#1}
\csname url@samestyle\endcsname
\providecommand{\newblock}{\relax}
\providecommand{\bibinfo}[2]{#2}
\providecommand{\BIBentrySTDinterwordspacing}{\spaceskip=0pt\relax}
\providecommand{\BIBentryALTinterwordstretchfactor}{4}
\providecommand{\BIBentryALTinterwordspacing}{\spaceskip=\fontdimen2\font plus
\BIBentryALTinterwordstretchfactor\fontdimen3\font minus
  \fontdimen4\font\relax}
\providecommand{\BIBforeignlanguage}[2]{{%
\expandafter\ifx\csname l@#1\endcsname\relax
\typeout{** WARNING: IEEEtranS.bst: No hyphenation pattern has been}%
\typeout{** loaded for the language `#1'. Using the pattern for}%
\typeout{** the default language instead.}%
\else
\language=\csname l@#1\endcsname
\fi
#2}}
\providecommand{\BIBdecl}{\relax}
\BIBdecl

\bibitem{Abadi2005}
M.~Abadi, M.~Budiu, U.~Erlingsson, and J.~Ligatti, ``Control-flow integrity,''
  in \emph{Proceedings of the 12th ACM Conference on Computer and
  Communications Security}, ser. CCS~'05, Alexandria, VA, USA, 2005, pp.
  340--353.

\bibitem{Avanzi2017:QARMA}
R.~Avanzi, ``The {QARMA} block cipher family. almost {MDS} matrices over rings
  with zero divisors, nearly symmetric even-mansour constructions with
  non-involutory central rounds, and search heuristics for low-latency
  {S}-boxes,'' \emph{IACR Transactions on Symmetric Cryptology}, vol. 2017,
  no.~1, pp. 4--44, Mar. 2017.

\bibitem{Berger2006:DieHard}
E.~D. Berger and B.~G. Zorn, ``Diehard: Probabilistic memory safety for unsafe
  languages,'' in \emph{Proceedings of the 27th ACM SIGPLAN Conference on
  Programming Language Design and Implementation}, ser. PLDI~'06, Ottawa,
  Ontario, Canada, 2006, pp. 158--168.

\bibitem{gem5}
N.~Binkert, B.~Beckmann, G.~Black, S.~K. Reinhardt, A.~Saidi, A.~Basu,
  J.~Hestness, D.~R. Hower, T.~Krishna, S.~Sardashti, R.~Sen, K.~Sewell,
  M.~Shoaib, N.~Vaish, M.~D. Hill, and D.~A. Wood, ``The {Gem5} simulator,''
  \emph{SIGARCH Computer Architecture News}, 2011.

\bibitem{Bittau2014:HackingBlind}
A.~Bittau, A.~Belay, A.~Mashtizadeh, D.~Mazi\`{e}res, and D.~Boneh, ``Hacking
  blind,'' in \emph{Proceedings of the 2014 IEEE Symposium on Security and
  Privacy}, ser. S\&P~'14, Washington, DC, USA, 2014, pp. 227--242.

\bibitem{Bletsch2011:JOP}
T.~Bletsch, X.~Jiang, V.~W. Freeh, and Z.~Liang, ``Jump-oriented programming: A
  new class of code-reuse attack,'' in \emph{Proceedings of the 6th ACM
  Symposium on Information, Computer and Communications Security}, ser.
  ASIACCS~'11, Hong Kong, China, 2011, pp. 30--40.

\bibitem{Flute}
Bluespec, ``Flute: 5-stage, in-order, piplined {RISC-V CPU},''
  \url{https://github.com/bluespec/Flute}, 2019, [Online; accessed
  15-June-2020].

\bibitem{Bounov2016}
D.~Bounov, R.~Gokhan~Kici, and S.~Lerner, ``Protecting {C++} dynamic dispatch
  through {VTable} interleaving,'' in \emph{Proceedings of the 2016 Network and
  Distributed System Security Symposium}, ser. NDSS~'16, San Diego, CA, USA,
  February 2016.

\bibitem{Bovet:2002:ULK}
D.~Bovet and M.~Cesati, \emph{Understanding the Linux Kernel, Second Edition},
  2nd~ed., A.~Oram, Ed.\hskip 1em plus 0.5em minus 0.4em\relax Sebastopol, CA,
  USA: O'Reilly \& Associates, Inc., 2002.

\bibitem{Bucek2018:SPEC}
J.~Bucek, K.-D. Lange, and J.~v.~Kistowski, ``{SPEC CPU2017}: Next-generation
  compute benchmark,'' in \emph{Proceedings of the 2018 ACM/SPEC International
  Conference on Performance Engineering}, ser. ICPE '18, Berlin, Germany, April
  2018, pp. 41--42.

\bibitem{Buchanan2008}
E.~Buchanan, R.~Roemer, H.~Shacham, and S.~Savage, ``When good instructions go
  bad: Generalizing return-oriented programming to {RISC},'' in
  \emph{Proceedings of the 15th ACM Conference on Computer and Communications
  Security}, ser. CCS~'08, Alexandria, Virginia, USA, 2008, pp. 27--38.

\bibitem{burow2017control}
N.~Burow, S.~A. Carr, J.~Nash, P.~Larsen, M.~Franz, S.~Brunthaler, and
  M.~Payer, ``Control-flow integrity: Precision, security, and performance,''
  \emph{ACM Computing Surveys (CSUR)}, vol.~50, no.~1, p.~16, 2017.

\bibitem{Burow2018}
N.~Burow, D.~McKee, S.~A.~Carr, and M.~Payer, ``{CFIXX}: Object type integrity
  for {C++} virtual dispatch,'' in \emph{Proceedings of the 2018 Network and
  Distributed System Security Symposium}, ser. NDSS~'18, San Diego, CA, USA,
  February 2018.

\bibitem{Burow2019:shadow}
N.~Burow, X.~Zhang, and M.~Payer, ``{SoK}: Shining light on shadow stacks,'' in
  \emph{Proceedings of the 2019 IEEE Symposium on Security and Privacy}, ser.
  S\&P~'19, May 2019.

\bibitem{Checkoway2010}
S.~Checkoway, L.~Davi, A.~Dmitrienko, A.-R. Sadeghi, H.~Shacham, and
  M.~Winandy, ``Return-oriented programming without returns,'' in
  \emph{Proceedings of the 17th ACM Conference on Computer and Communications
  Security}, ser. CCS~'10, Chicago, Illinois, USA, 2010, pp. 559--572.

\bibitem{Cheng2014}
Y.~Cheng, Z.~Zhou, M.~Yu, X.~Ding, and R.~H. Deng, ``Ropecker: A generic and
  practical approach for defending against {ROP} attacks,'' in
  \emph{Proceedings of the 2014 Network and Distributed System Security
  Symposium}, ser. NDSS~'16, San Diego, CA, USA, February 2014.

\bibitem{cowan1998stackguard}
C.~Cowan, C.~Pu, D.~Maier, J.~Walpole, P.~Bakke, S.~Beattie, A.~Grier,
  P.~Wagle, Q.~Zhang, and H.~Hinton, ``Stackguard: Automatic adaptive detection
  and prevention of buffer-overflow attacks.'' in \emph{USENIX security
  symposium}, vol.~98, 1998, pp. 63--78.

\bibitem{Cowan2003:PointGuard}
C.~Cowan, S.~Beattie, J.~Johansen, and P.~Wagle, ``Pointguard: Protecting
  pointers from buffer overflow vulnerabilities,'' in \emph{Proceedings of the
  12th Conference on USENIX Security Symposium - Volume 12}, ser. SSYM~'03,
  Washington, DC, USA, 2003, p.~7.

\bibitem{Cox2006}
B.~Cox, D.~Evans, A.~Filipi, J.~Rowanhill, W.~Hu, J.~Davidson, J.~Knight,
  A.~Nguyen-Tuong, and J.~Hiser, ``N-variant systems: A secretless framework
  for security through diversity,'' in \emph{Proceedings of the 15th Conference
  on USENIX Security Symposium - Volume 15}, ser. USENIX-SS'06, Vancouver,
  B.C., Canada, 2006.

\bibitem{Dahl1972:StructuredProg}
O.-J. Dahl, E.~W. Dijkstra, and C.~A.~R. Hoare, Eds., \emph{Structured
  Programming}.\hskip 1em plus 0.5em minus 0.4em\relax London, UK: Academic
  Press Ltd., 1972.

\bibitem{isomeron2015}
L.~Davi, C.~Liebchen, A.-R. Sadeghi, K.~Z.~Snow, and F.~Monrose, ``Isomeron:
  Code randomization resilient to ({Just-In-Time}) return-oriented
  programming,'' in \emph{Proceedings of the 2015 Network and Distributed
  System Security Symposium}, ser. NDSS~'15, San Diego, CA, USA, February 2015.

\bibitem{Davi2013:GadgeMe}
L.~V. Davi, A.~Dmitrienko, S.~N\"{u}rnberger, and A.-R. Sadeghi, ``Gadge me if
  you can: Secure and efficient ad-hoc instruction-level randomization for
  {x86} and {ARM},'' in \emph{Proceedings of the 8th ACM SIGSAC Symposium on
  Information, Computer and Communications Security}, ser. ASIA CCS~'13,
  Hangzhou, China, 2013, pp. 299--310.

\bibitem{drepper2004security}
U.~Drepper, ``Security enhancements in redhat enterprise {Linux} (beside
  {SELinux}),'' 2004.

\bibitem{exploitDB:Apache}
{Exploit Database}, ``Apache 2.4.7 + php 7.0.2 - openssl\_seal() uninitialized
  memory code execution,'' \url{https://www.exploit-db.com/exploits/40142},
  [Online; accessed 15-June-2020].

\bibitem{exploitDB:mcrypt}
------, ``mcrypt 2.5.8 - local stack overflow,''
  \url{https://www.exploit-db.com/exploits/22928}, [Online; accessed
  15-June-2020].

\bibitem{exploitDB:Netperf}
------, ``Netperf 2.6.0 - stack-based buffer overflow,''
  \url{https://www.exploit-db.com/exploits/46997}, [Online; accessed
  15-June-2020].

\bibitem{exploitDB:Nginx}
------, ``Nginx 1.3.9 < 1.4.0 - chuncked encoding stack buffer overflow,''
  \url{https://www.exploit-db.com/exploits/25775}, [Online; accessed
  15-June-2020].

\bibitem{Gallagher2019:Morpheus}
M.~Gallagher, L.~Biernacki, S.~Chen, Z.~B. Aweke, S.~F. Yitbarek, M.~T. Aga,
  A.~Harris, Z.~Xu, B.~Kasikci, V.~Bertacco, S.~Malik, M.~Tiwari, and
  T.~Austin, ``Morpheus: A vulnerability-tolerant secure architecture based on
  ensembles of moving target defenses with churn,'' in \emph{Proceedings of the
  Twenty-Fourth International Conference on Architectural Support for
  Programming Languages and Operating Systems}, ser. ASPLOS~'19, Providence,
  RI, USA, 2019, pp. 469--484.

\bibitem{Gawlik2016:Detile}
R.~Gawlik, P.~Koppe, B.~Kollenda, A.~Pawlowski, B.~Garmany, and T.~Holz,
  ``Detile: Fine-grained information leak detection in script engines,'' in
  \emph{Proceedings of the 13th International Conference on Detection of
  Intrusions and Malware, and Vulnerability Assessment - Volume 9721}, ser.
  DIMVA~'16, San Sebastian, Spain, 2016, pp. 322--342.

\bibitem{goktas2014out}
E.~G{\"o}ktas, E.~Athanasopoulos, H.~Bos, and G.~Portokalidis, ``Out of
  control: Overcoming control-flow integrity,'' in \emph{Proceedings of the
  2014 IEEE Symposium on Security and Privacy}, ser. S\&P~'14, San Jose, CA,
  USA, may 2014, pp. 575--589.

\bibitem{goktas2016:InfoHiding}
E.~G{\"o}kta{\c s}, R.~Gawlik, B.~Kollenda, E.~Athanasopoulos, G.~Portokalidis,
  C.~Giuffrida, and H.~Bos, ``Undermining information hiding (and what to do
  about it),'' in \emph{Proceedings of the 25th {USENIX} Security Symposium
  ({USENIX} Security 16)}, Austin, TX, Aug. 2016, pp. 105--119.

\bibitem{Ha2018}
D.~{Ha}, W.~{Jin}, and H.~{Oh}, ``{REPICA}: Rewriting position independent code
  of {ARM},'' \emph{IEEE Access}, vol.~6, pp. 50\,488--50\,509, 2018.

\bibitem{Hu2016}
H.~{Hu}, S.~{Shinde}, S.~{Adrian}, Z.~L. {Chua}, P.~{Saxena}, and Z.~{Liang},
  ``Data-oriented programming: On the expressiveness of non-control data
  attacks,'' in \emph{Proceedings of the 2016 IEEE Symposium on Security and
  Privacy}, ser. S\&P~'16, San Jose, CA, USA, May 2016, pp. 969--986.

\bibitem{IntelCeT}
Intel, ``Intel control-flow enforcement technology preview,''
  \url{https://software.intel.com/sites/default/files/managed/4d/2a/control-flow-enforcement-technology-preview.pdf},
  2017, [Online; accessed 15-June-2020].

\bibitem{Ispoglou2018:BOP}
K.~K. Ispoglou, B.~AlBassam, T.~Jaeger, and M.~Payer, ``Block oriented
  programming: Automating data-only attacks,'' in \emph{Proceedings of the 2018
  ACM SIGSAC Conference on Computer and Communications Security}, ser. CCS~'18,
  Toronto, Canada, 2018, pp. 1868--1882.

\bibitem{Gaurav2003:ISR}
G.~S. Kc, A.~D. Keromytis, and V.~Prevelakis, ``Countering code-injection
  attacks with instruction-set randomization,'' in \emph{Proceedings of the
  10th ACM Conference on Computer and Communications Security}, ser. CCS~'03,
  Washington D.C., USA, 2003, pp. 272--280.

\bibitem{Kim90:meta}
L.-S. Kim and R.~W. Dutton, ``Metastability of {CMOS} latch/flip-flop,''
  \emph{IEEE Journal of Solid-State Circuits}, vol.~25, no.~4, pp. 942--951,
  Aug 1990.

\bibitem{Kim2017}
T.~Kim, C.~H. Kim, H.~Choi, Y.~Kwon, B.~Saltaformaggio, X.~Zhang, and D.~Xu,
  ``{RevARM}: A platform-agnostic {ARM} binary rewriter for security
  applications,'' in \emph{Proceedings of the 33rd Annual Computer Security
  Applications Conference}, ser. ACSAC~'17, Orlando, FL, USA, 2017, pp.
  412--424.

\bibitem{Kim2014}
Y.~Kim, R.~Daly, J.~Kim, C.~Fallin, J.~H. Lee, D.~Lee, C.~Wilkerson, K.~Lai,
  and O.~Mutlu, ``Flipping bits in memory without accessing them: An
  experimental study of {DRAM} disturbance errors,'' in \emph{Proceeding of the
  41st Annual International Symposium on Computer Architecuture}, ser.
  ISCA~'14, Minneapolis, Minnesota, USA, 2014, pp. 361--372.

\bibitem{Koning2016}
K.~{Koning}, H.~{Bos}, and C.~{Giuffrida}, ``Secure and efficient multi-variant
  execution using hardware-assisted process virtualization,'' in
  \emph{Proceedings of the 2016 46th Annual IEEE/IFIP International Conference
  on Dependable Systems and Networks}, ser. DSN~'16, June 2016, pp. 431--442.

\bibitem{Kwon2013:LowFatHW}
A.~Kwon, U.~Dhawan, J.~M. Smith, T.~F. Knight, Jr., and A.~DeHon, ``Low-fat
  pointers: Compact encoding and efficient gate-level implementation of fat
  pointers for spatial safety and capability-based security,'' in
  \emph{Proceedings of the 2013 ACM SIGSAC Conference on Computer \&
  Communications Security}, ser. CCS~'13, Berlin, Germany, 2013, pp. 721--732.

\bibitem{Kwon2016:LDX}
Y.~Kwon, D.~Kim, W.~N. Sumner, K.~Kim, B.~Saltaformaggio, X.~Zhang, and D.~Xu,
  ``{LDX}: Causality inference by lightweight dual execution,'' in
  \emph{Proceedings of the Twenty-First International Conference on
  Architectural Support for Programming Languages and Operating Systems}, ser.
  ASPLOS~'16, Atlanta, Georgia, USA, 2016, pp. 503--515.

\bibitem{Hans2018}
H.~Liljestrand, T.~Nyman, K.~Wang, C.~C. Perez, J.-E. Ekberg, and N.~Asokan,
  ``{PAC} it up: Towards pointer integrity using {ARM} pointer
  authentication,'' in \emph{Proceedings of the 28th {USENIX} Security
  Symposium}, ser. ({USENIX} Security 19), Santa Clara, CA, USA, Aug. 2019, pp.
  177--194.

\bibitem{llvm-cfi}
LLVM, ``Control flow integrity design,''
  \url{https://clang.llvm.org/docs/ControlFlowIntegrityDesign.html}, [Online;
  accessed 15-June-2020].

\bibitem{Locasto2005:AC}
M.~E. Locasto, S.~Sidiroglou, and A.~D. Keromytis, ``Application communities:
  Using monoculture for dependability,'' in \emph{Proceedings of the First
  Conference on Hot Topics in System Dependability}, ser. HotDep’05.\hskip
  1em plus 0.5em minus 0.4em\relax Yokohama, Japan: USENIX Association, 2005,
  p.~9.

\bibitem{Lu2018:Buddy}
K.~{Lu}, M.~{Xu}, C.~{Song}, T.~{Kim}, and W.~{Lee}, ``Stopping memory
  disclosures via diversification and replicated execution,'' \emph{IEEE
  Transactions on Dependable and Secure Computing}, 2018.

\bibitem{Mashtizadeh2015}
A.~J. Mashtizadeh, A.~Bittau, D.~Boneh, and D.~Mazi\`{e}res, ``{CCFI}:
  Cryptographically enforced control flow integrity,'' in \emph{Proceedings of
  the 22Nd ACM SIGSAC Conference on Computer and Communications Security}, ser.
  CCS~'15, Denver, Colorado, USA, 2015, pp. 941--951.

\bibitem{musl}
musl, ``musl libc,'' \url{http://www.musl-libc.org/}, [Online; accessed
  15-June-2020].

\bibitem{Nergal2011}
Nergal, ``The advanced return-into-lib(c) exploits: {PaX} case study,''
  \url{http://phrack.org/issues/58/4.html}, 2001, [Online; accessed
  15-June-2020].

\bibitem{Nyman2017}
T.~Nyman, J.-E. Ekberg, L.~Davi, and N.~Asokan, ``{CFI CaRE}:
  Hardware-supported call and return enforcement for commercial
  microcontrollers,'' in \emph{Research in Attacks, Intrusions, and
  Defenses}.\hskip 1em plus 0.5em minus 0.4em\relax Springer International
  Publishing, 2017, pp. 259--284.

\bibitem{Antonis2013:ASIST}
A.~Papadogiannakis, L.~Loutsis, V.~Papaefstathiou, and S.~Ioannidis, ``{ASIST}:
  Architectural support for instruction set randomization,'' in
  \emph{Proceedings of the 2013 ACM SIGSAC Conference on Computer and
  Communications Security}, ser. CCS~'13, Berlin, Germany, 2013, pp. 981--992.

\bibitem{smash2012}
V.~{Pappas}, M.~{Polychronakis}, and A.~D. {Keromytis}, ``Smashing the gadgets:
  Hindering return-oriented programming using in-place code randomization,'' in
  \emph{Proceedings of the 2012 IEEE Symposium on Security and Privacy}, ser.
  S\&P~'12, May 2012, pp. 601--615.

\bibitem{Qualcomm2017}
I.~Qualcomm~Technologies, ``Pointer authentication on {ARMv8.3},''
  \url{https://www.qualcomm.com/media/documents/files/whitepaper-pointer-authentication-on-armv8-3.pdf},
  2017, [Online; accessed 15-June-2020].

\bibitem{Sasaki2019:Califorms}
H.~Sasaki, M.~A. Arroyo, M.~T.~I. Ziad, K.~Bhat, K.~Sinha, and
  S.~Sethumadhavan, ``Practical byte-granular memory blacklisting using
  {C}aliforms,'' in \emph{Proceedings of the 52nd Annual IEEE/ACM International
  Symposium on Microarchitecture}, ser. MICRO~'52, Columbus, OH, USA, 2019, pp.
  558--571.

\bibitem{Ropper}
S.~Schirra, ``Ropper,'' \url{https://github.com/sashs/Ropper}, [Online;
  accessed 15-June-2020].

\bibitem{Schuster2015}
F.~Schuster, T.~Tendyck, C.~Liebchen, L.~Davi, A.-R. Sadeghi, and T.~Holz,
  ``Counterfeit object-oriented programming: On the difficulty of preventing
  code reuse attacks in {C++} applications,'' in \emph{Proceedings of the 2015
  IEEE Symposium on Security and Privacy}, ser. S\&P~'15, Oakland, CA, USA,
  2015, pp. 745--762.

\bibitem{Shacham2007}
H.~Shacham, ``The geometry of innocent flesh on the bone: Return-into-libc
  without function calls (on the x86),'' in \emph{Proceedings of the 14th ACM
  Conference on Computer and Communications Security}, ser. CCS~'07,
  Alexandria, Virginia, USA, 2007, pp. 552--561.

\bibitem{Shanbhogue2019:IntelCeTPaper}
V.~Shanbhogue, D.~Gupta, and R.~Sahita, ``Security analysis of processor
  instruction set architecture for enforcing control-flow integrity,'' in
  \emph{Proceedings of the 8th International Workshop on Hardware and
  Architectural Support for Security and Privacy}, ser. HASP '19, Phoenix, AZ,
  USA, 2019.

\bibitem{Sinha2017:ISR}
K.~Sinha, V.~P. Kemerlis, and S.~Sethumadhavan, ``Reviving instruction set
  randomization,'' in \emph{Proceedings of the 2017 IEEE International
  Symposium on Hardware Oriented Security and Trust (HOST)}, 2017, pp. 21--28.

\bibitem{Kanad2018:REST}
K.~Sinha and S.~Sethumadhavan, ``Practical memory safety with {REST},'' in
  \emph{Proceedings of the 45th Annual International Symposium on Computer
  Architecture}, ser. ISCA~'18, Los Angeles, California, USA, 2018, pp.
  600--611.

\bibitem{snow2013}
K.~Z. {Snow}, F.~{Monrose}, L.~{Davi}, A.~{Dmitrienko}, C.~{Liebchen}, and
  A.-R. {Sadeghi}, ``Just-in-time code reuse: On the effectiveness of
  fine-grained address space layout randomization,'' in \emph{Proceedings of
  the 2013 IEEE Symposium on Security and Privacy}, ser. S\&P~'13, Berkeley,
  CA, USA, May 2013, pp. 574--588.

\bibitem{ret2libc_solar:1997}
{Solar Designer}, ``Getting around non-executable stack (and fix),''
  \url{http://seclists.org/bugtraq/1997/Aug/63}, August 1997.

\bibitem{Statista:Servers}
Statista, Gartner, and IDC, ``Server shipments worldwide from 2010 to 2018.''
  \url{https://www.statista.com/statistics/219596/worldwide-server-shipments-by-vendor/},
  2019, [Online; accessed 15-June-2020].

\bibitem{Statista:MCU}
Statista and I.~Insights, ``Microcontroller unit (mcu) shipments worldwide from
  2015 to 2023.''
  \url{https://www.statista.com/statistics/935382/worldwide-microcontroller-unit-shipments/},
  2019, [Online; accessed 15-June-2020].

\bibitem{Szekeres2013}
L.~Szekeres, M.~Payer, T.~Wei, and D.~Song, ``{SoK}: Eternal war in memory,''
  in \emph{Proceedings of the 2013 IEEE Symposium on Security and Privacy},
  ser. S\&P~'13, San Francisco, CA, USA, 2013, pp. 48--62.

\bibitem{Tuck2004:PointerEncryption}
N.~{Tuck}, B.~{Calder}, and G.~{Varghese}, ``Hardware and binary modification
  support for code pointer protection from buffer overflow,'' in \emph{37th
  International Symposium on Microarchitecture (MICRO~'04)}, Portland, OR, USA,
  2004, pp. 209--220.

\bibitem{Volckaert2016:DCL}
S.~{Volckaert}, B.~{Coppens}, and B.~{De Sutter}, ``Cloning your gadgets:
  Complete {ROP} attack immunity with multi-variant execution,'' \emph{IEEE
  Transactions on Dependable and Secure Computing}, vol.~13, no.~4, pp.
  437--450, July 2016.

\bibitem{Volckaert2016}
S.~Volckaert, B.~Coppens, A.~Voulimeneas, A.~Homescu, P.~Larsen, B.~De~Sutter,
  and M.~Franz, ``Secure and efficient application monitoring and
  replication,'' in \emph{Proceedings of the 2016 USENIX Conference on Usenix
  Annual Technical Conference}, ser. ATC~'16, Denver, CO, USA, 2016, pp.
  167--179.

\bibitem{Watson2015:CHERI}
R.~N.~M. Watson, J.~Woodruff, P.~G. Neumann, S.~W. Moore, J.~Anderson,
  D.~Chisnall, N.~H. Dave, B.~Davis, K.~Gudka, B.~Laurie, S.~J. Murdoch, R.~M.
  Norton, M.~Roe, S.~D. Son, and M.~Vadera, ``{CHERI}: A hybrid
  capability-system architecture for scalable software compartmentalization,''
  in \emph{Proceedings of the 2015 IEEE Symposium on Security and Privacy}, San
  Jose, CA, USA, May 2015, pp. 20--37.

\bibitem{Ollie2007}
O.~Whitehouse, ``An analysis of address space layout randomization on windows
  vista,'' Jan 2007.

\bibitem{Wilander2011:RIPE}
J.~Wilander, N.~Nikiforakis, Y.~Younan, M.~Kamkar, and W.~Joosen, ``{RIPE}:
  Runtime intrusion prevention evaluator,'' in \emph{Proceedings of the 27th
  Annual Computer Security Applications Conference}, ser. ACSAC '11, Orlando,
  Florida, USA, 2011, pp. 41--50.

\bibitem{David2016:shuffler}
D.~Williams-King, G.~Gobieski, K.~Williams-King, J.~P. Blake, X.~Yuan, P.~Colp,
  M.~Zheng, V.~P. Kemerlis, J.~Yang, and W.~Aiello, ``Shuffler: Fast and
  deployable continuous code re-randomization,'' in \emph{Proceedings of the
  12th {USENIX} Symposium on Operating Systems Design and Implementation}, ser.
  OSDI~16.\hskip 1em plus 0.5em minus 0.4em\relax Savannah, GA, USA: USENIX
  Association, 2016, pp. 367--382.

\bibitem{williams2020egalito}
D.~Williams-King, H.~Kobayashi, K.~Williams-King, G.~Patterson, F.~Spano, Y.~J.
  Wu, J.~Yang, and V.~P. Kemerlis, ``Egalito: Layout-agnostic binary
  recompilation,'' in \emph{Proceedings of the Twenty-Fifth International
  Conference on Architectural Support for Programming Languages and Operating
  Systems}, 2020, pp. 133--147.

\bibitem{Chao2016}
C.~Zhang, S.~A.~Carr, T.~Li, Y.~Ding, C.~Song, M.~Payer, and D.~Song, ``Vtrust:
  Regaining trust on virtual calls,'' in \emph{Proceedings of the 2016 Network
  and Distributed System Security Symposium}, ser. NDSS~'16, San Diego, CA,
  USA, February 2016.

\end{thebibliography}
}

\end{document}